\documentclass[reprint, showpacs, showkeys, nofootinbib, aps, prd]{revtex4-1}
\usepackage{bm}
\usepackage{amsmath}
\usepackage{mathptmx}
\usepackage{amsfonts}
\usepackage{amssymb}
\usepackage[pdftex]{graphicx}
\usepackage{color}
\usepackage{mathptmx}
\RequirePackage[colorlinks,citecolor=darkred,urlcolor=magenta,linkcolor=darkblue,bookmarks=false]{hyperref}
\usepackage{subfig}
\usepackage[toc,title]{appendix}
\usepackage[misc]{ifsym}
\usepackage{adjustbox}
\usepackage{multirow}
\usepackage{makecell}

\allowdisplaybreaks

\definecolor{darkblue}{rgb}{0.0, 0.0, 0.62}
\definecolor{deepmagenta}{rgb}{0.8, 0.0, 0.7}
\definecolor{darkred}{rgb}{0.55, 0.0, 0.0}

\RequirePackage{./mymacro}
   \graphicspath{{./graphs/}}

\begin{document}

\title{Differentiating dark interactions with perturbation}
\author{Srijita Sinha}
\email[\Letter \hspace{0.1cm}]{ss13ip012@iiserkol.ac.in}
\affiliation{Indian Institute of Science Education and Research Kolkata, Mohanpur, Nadia, 741246, India}%

\begin{abstract}
A cosmological model with an energy transfer between dark matter (DM) and dark energy (DE) can give rise to comparable energy densities at the present epoch. The present work deals with the perturbation analysis, parameter estimation and Bayesian evidence calculation of interacting models with dynamical coupling parameter that determines the strength of the interaction. We have considered two cases, where the interaction is a more recent phenomenon and where the interaction is a phenomenon in the distant past. Moreover, we have considered the quintessence DE equation of state with Chevallier-Polarski-Linder (CPL) parametrisation and energy flow from DM to DE. Using the current observational datasets like the cosmic microwave background (CMB), baryon acoustic oscillation (BAO), Type Ia Supernovae (SNe Ia) and redshift-space distortions (RSD), we have estimated the mean values of the parameters. Using the perturbation analysis and Bayesian evidence calculation, we have shown that interaction present as a brief early phenomenon is preferred over a recent interaction.
\end{abstract}


\maketitle
\section{Introduction} \label{sec:intro}
The discovery that the Universe is expanding with an acceleration~\cite{riess1998anj, schmidt1998apj,perlmutter1999apj,scolnic2018apj} has set a new milestone in the history of cosmology. This discovery also presented a new challenge as explaining this phenomenon requires an agent that leads to a repulsive gravity. The flurry of more recent high precision observational data~\cite{eisenstein1998apj, planck2018cp, partphys2018prd, reid2010mnras, alam2017sdss3, descol2018prd1, descol2018prd2, descol2019prl, alam2020sdss4} has consolidated the fact that the Universe indeed gravitates in the wrong way. Many theoretical models have been put forward to explain the repulsive nature of gravity, but arguably the most popularly accepted one is the presence of an exotic component named ``dark energy''. This exotic component of the contents of the Universe can produce a sufficient negative pressure, which overcomes the gravitational attraction of matter and drives the recent acceleration. The cosmological constant, $\Lambda$~\cite{paddy2003pr, copeland2006ijmpd, amendola2010prl, wands2012cqg, mehrabi2018prd, planck2018cp, martinelli2019mnras} is the first preferred choice, followed by a scalar field with a potential~\cite{frieman1995prl,carroll1998prl, caldwell1998prl, sahni2000ijmpd, urena2000prd, carroll2001lrr,  peebles2003rmp, copeland2005prd, sinha2021jcap}. The other popular choices include Holographic dark energy~\cite{li2004plb, pavon2006aip, zimdahl2007cqg, elizalde2004prd, nojiri2006grg, zhang2012jcap, chimento2012prd, akhlaghi2018mnras}, Chaplygin gas~\cite{kamenshchik2001plb, bilic2002plb, bento2002prd, padmanabhan2002prd, chimento2011prd, wang2013prd}, phantom field~\cite{caldwell2002plb, carroll2003prd}, quintom model~\cite{feng2005plb,cai2007plb1}, to name a few. The list of candidates as dark energy is far from being complete in the absence of a universally accepted one. There are excellent reviews~\cite{sahni2006ijmpd, tsujikawa2013cqg, sami2016ijmpd, brax2018rpp} on these candidates. 

The observationally most preferred model, $\Lambda$ with cold dark matter (\lcdm), faces many problems like the so-called ``cosmological constant problem''~\cite{sahni2000ijmpd, sahni2002cqg} and the coincidence problem~\cite{steinhardt2003jstor,velten2014epjc}. The cosmological constant problem is the discrepancy between the theoretical value and the observed value of the cosmological constant. The coincidence problem is the question, why both dark matter and dark energy have comparable energy densities precisely at the present epoch? These problems in the simple \lcdm model are the motivation to search for other possible avenues.

The fact that dark matter and dark energy have energy densities of the same order of magnitude opens the possibility that there is an energy exchange between the two. Interactions between dark matter and dark energy in various dark energy models have been studied and tested against observations extensively~\cite{billyard2000prd, pavon2004jcap, amendola2004jcap, curbelo2006cqg, gonzalez2006cqg,guo2007prd,olivares2008prd, bohmer2008prd, quercellini2008prd, bean2008prd2, quartin2008jcap, he2008jcap, chimento2010prd, amendola2012prd, pettorino2012prd, salvatelli2014prl, yang2014prd1, wang2014aa, caprini2016jcap, nunes2016prd, mukherjee2017cqg, yang2017prd2, pan2018mnras, yang2018prd, yang2018jcap, visinelli2019prd, vagnozzi2020mnras}. For detailed reviews on interacting dark matter-dark energy models, we refer to~\cite{bamba2012ass, bolotin2015ijmpd, wang2016rpp}.

The presence of a coupling in the dark sector may not be ruled out \emph{a priori}~\cite{billyard2000prd, pavon2004jcap, amendola2004jcap, curbelo2006cqg, gonzalez2006cqg, guo2007prd, olivares2008prd, bohmer2008prd, quercellini2008prd, bean2008prd2, quartin2008jcap, he2008jcap, caldera2009jcap, chimento2010prd, amendola2012prd, pettorino2012prd, chimento2012prd2, chimento2013prd, salvatelli2014prl, yang2014prd1, wang2014aa, caprini2016jcap, nunes2016prd, mukherjee2017cqg, yang2017prd2, pan2018mnras, yang2018prd, yang2018mnras, bruck2017prd}. It naturally raises the question whether the interaction was there from the beginning of the Universe and exists through its evolution or is a recent phenomenon, or it was entirely an early phenomenon and not at all present today. A modification of the phenomenological interaction term by an evolving coupling parameter instead of its being a constant, may answer this question. A constant coupling parameter indicates the interaction is present throughout the evolution of the Universe~\cite{guo2007prd, yang2017prd1}. In this work, we have considered the coupling parameter to be evolving with the scale factor. Interaction with an evolving coupling parameter is not studied much in literature and warrants a detailed analysis. Rosenfeld~\cite{rosenfeld2007prd} and Yang \etal~\cite{yang2019du} have considered the dynamical coupling parameter, but the motivation as well as the analytical form of the parameter used in the present work are different.

There is no theoretically preferred form of the phenomenological interaction term. In this work, two possible scenarios are considered --- (a) the presence of interaction is significant during the late time but not at early time and (b) the presence of interaction is significant in the early times but not at late time. The rate of energy transfer is considered to be proportional to the dark energy density. The dynamical coupling parameter will affect the evolution of the dark matter and hence have its imprints in the growth of perturbations. Thus the presence of dynamical interaction can give rise to new features in structure formation. The motivation of the present work is to investigate the effect of interactions on clustering of matter perturbation and test the models against observational datasets. 

We have tested the interacting models with different observational datasets like the cosmic microwave background (CMB)~\cite{planck2018cp}, baryon acoustic oscillation (BAO)~\cite{beutler2011mnras, ross2015mnras, alam2017sdss3}, Type Ia Supernovae (SNe Ia)~\cite{scolnic2018apj} data and their different combinations. For a complete understanding of the effect of interaction on structure formation, it is necessary to consider the effect of the large scale structure (LSS) information on the cosmological constraints. In the present work, we have considered the redshift-space distortions (RSD) data~\cite{kaiser1987mnras} as the LSS data. Combining the RSD data with CMB, BAO and Supernovae data is expected to break the degeneracy between the different interacting models with similar background evolution as well as provide a tight constraint on the interaction parameter.

The LSS data, which includes Planck Sunyaev-Zel'dovich survey~\cite{planck2013szc}, Canada France Hawaii Telescope Lensing Survey (CFHTLens)~\cite{kilbinger2013mnras,heymans2013mnras}, South Pole Telescope (SPT)~\cite{schaffer2011apj,vanengelen2012apj}, RSD survey, are in disagreement with CMB observations for the root-mean-square mass fluctuation in spheres with radius $8 h^{-1}\, \mpc$, (called $\se$) and hence for the matter density parameter $\Omega_{m}$ and the Hubble parameter $H_{0}$~\cite{pourtsidou2016prd, vandebruck2018prd, mohanty2018jaa, an2018jcap, martinelli2019mnras, lambiase2019epjc, eoin2019plb, banerjee2020prdl}. The LSS observations prefer lower values of $\se$ and $\Omega_{m}$ and a higher value of $H_{0}$ compared to the CMB results. Many attempts have been made to settle the disagreement between the two datasets~\cite{gomez2017epl, sakr2018aa, kazantzidis2018prd, gomez2018mnras, ooba2019ass, park2020prd}. Some more of the notable work with RSD data are~\cite{wang2014prd, yang2014prd1, yang2014prd2, costa2017jcap, nesseris2017prd, akhlaghi2018mnras, sagredo2018prd, skara2020prd, borges2020prd}.

The most persisting tension in observational cosmology is the discrepancy in the value of the Hubble parameter, $H_{0}$, as provided by the CMB measurement from the \Planck satellite and the local measurements like the Supernovae and $H_{0}$ for the Equation of State (SH0ES) project~\cite{riess2011apj, riess2018apj, riess2019apj}. The distance-ladder estimate of $H_{0} = 74.03\pm1.42$ $\mbox{km s}^{-1} \mbox{Mpc}^{-1}$ from the latest Hubble Space Telescope (HST) data~\cite{riess2019apj} increases the tension with the recent CMB measurement of $H_{0} = 67.36\pm 0.54$ $\mbox{km s}^{-1} \mbox{Mpc}^{-1}$~\cite{planck2018cp} to $4.4\sigma$. Other distance-ladder probes like the LIGO~\cite{abbott2017nature}, H0LiCOW~\cite{birrer2019mnras} do not seem to relieve the tension. The $H_{0}$ tension is more severe than the $\se$ tension. The $\se$ and $H_{0}$ tensions can be attributed to the possible systematics in the CMB or local measurements~\cite{aghanim2017aa, planck2019cmb, jones2018aas, rigault2020aa}. On the other hand, these tensions strengthen the reason to search for models other than the simple \lcdm model. Inspite of the many attempts~\cite{clifton2012pr, divalentino2016plb, bernal2016jcap, divalentino2016prd, ezquiaga2017prl, alam2017prd, divalentino2017prd1, divalentino2017prd2, frusciante2020pr} towards the resolution, the tension still persists. For detail review on $H_{0}$ tension, we refer to~\cite{jackson2007lrr, verde2019na}. A non-gravitational interaction between the dark components is often introduced to attempt a resolution to the tension with some success. The simplest interacting model without introducing any new degrees of freedom is the interaction of dark matter with the inhomogeneous vacuum energy density as shown in~\cite{wands2012cqg, alcaniz2012plb, wang2013prd, salvatelli2014prl, velten2015mnras, marttens2017pdu, kumar2017prd, martinelli2019mnras, kumar2019epjc, borges2020prd, hogg2020pdu, divalentino2020pdu}. Many other interacting dark energy models have beed studied throughly in the literature~\cite{battye2014prl, yang2014prd1, divalentino2015prd, divalentino2017prd1, divalentino2017prd2, yang2018jcap, divalentino2018sym, pandey2020jcap, vagnozzi2020prd, divalentino2020pdu, divalentino2020prd, yang2020mnras, yang2020prd}.  

It must be mentioned here that the model with constant coupling parameter has been tested rigorously against different observational datasets and priors ranges~\cite{martinelli2019mnras, divalentino2020pdu, vagnozzi2020prd} to name a few. In this work, we used different datasets and different prior ranges and an `evolving' coupling parameter in the interaction term. Moreover, we considered an evolving dark energy with EoS given by the Chevallier-Polarski-Linder (CPL) parametrisation. However, the present work is not an attempt to alleviate the $\se$ or $H_{0}$ tensions but to understand the evolution of the interaction using perturbation and test the models against observational datasets.

The paper is organised as follows. Section \ref{sec:bckgrnd} discusses the background equations of interacting dark matter-dark energy models. The perturbation equations, evolution of the density contrast along with the effects on the cosmic microwave background (CMB) temperature fluctuation, matter power spectrum, linear growth rate and $\fsg$ are discussed in Sect.\ \ref{sec:pert}. In Sect.\ \ref{sec:mcmc}, we discuss the results obtained from constraining the interacting models against different observational datasets performing the Markov Chain Monte Carlo (MCMC) analysis and in Sect.\ \ref{sec:evi}, we discuss our inference from Bayesian evidence calculation. Finally, in Sect.\ \ref{sec:sum}, we conclude with a summary and a brief discussion of the results that we arrived at. The details on the datasets used and the method are given in Appendices \ref{sec:app} and \ref{sec:model-selection}.

\section{Interacting dark matter-dark energy fluid} \label{sec:bckgrnd}
The Universe is considered to be described by a spatially flat, homogeneous and isotropic Friedmann-Lema\^itre-Robertson-Walker (FLRW) metric,
\begin{equation}\label{eq:metric}
ds^2= a^2(\tau)\paren*{- d \tau ^2+\delta_{ij} d x^i dx^j},
\end{equation}
where $a(\tau)$ is the conformal scale factor and the relation between conformal time ($\tau$) and cosmic time ($t$) is $a^2 d\tau^2 = dt^2$.  Using the metric (Eqn.\ (\ref{eq:metric})), the Friedmann equations are written as
\begin{eqnarray}
3 \cH^2 &=& -a^2 \kappa \sum_{A}\rA \label{eq:fd1},\\ 
\cH^2+ 2 \cH^\prime &=& a^2 \kappa \sum_{A} \pA , \label{eq:fd2}
\end{eqnarray}
where $\kappa=8 \pi G_N$ ($G_N$ being the Newtonian Gravitational constant), $\cH\paren*{\tau}= \frac{a^\prime}{a}$ is the Hubble parameter and $\rA$ and $\pA$ are respectively the energy density and pressure of the different components of the Universe. A prime indicates differentiation with respect to the conformal time $\tau$. The Universe is filled with five components of matter, all formally represented as perfect fluids --- photons ($\gamma$), neutrinos ($\nu$), baryons ($b$), cold dark matter ($c$) and dark energy ($de$). We assume that there is an energy transfer only in the dark sector of the Universe such that the conservation equations are
\begin{eqnarray}
\rho^\prime_c+ 3 \cH \rho_c &=& -aQ\,,\label{eq:con1}\\
\rho^\prime_{de}+ 3 \cH \paren*{1+\wde} \rde&=& aQ. \label{eq:con2}
\end{eqnarray}
The pressure, $p_{c} = 0$ for cold dark matter. The other three fluids --- photons ($\gamma$), neutrinos ($\nu$) and baryons ($b$) conserve independently and hence, have no energy transfer among them. Their conservation equations are written as
\begin{equation}
\rho^\prime_{A}+ 3 \cH \paren*{1+w_{A}} \rA = 0\, \label{eq:con3},
\end{equation}
where $w_{A} = \pA/\rA$ is the equation of state parameter (EoS) of the $A$-th fluid and $A = \gamma, \nu, b$. For photons and neutrinos, the EoS parameter is $w_{\gamma} = w_{\nu} = 1/3$, for baryons and cold dark matter, the EoS parameter is $w_{b} = w_{c} = 0$ and for dark energy, the EoS parameter is $\wde = \pde/\rde$. 

In Eqns.\ (\ref{eq:con1}) and (\ref{eq:con2}), $Q$ gives the rate of energy transfer between the two fluids. If $Q<0$, energy is transferred from dark energy to dark matter (DE $\rightarrow$ DM) and if $Q>0$, energy is transferred from dark matter to dark energy (DM $\rightarrow$ DE). When $Q>0$, dark matter redshifts faster than $a^{-3}$ and when $Q<0$, dark matter redshifts slower than $a^{-3}$. The dark energy evolution depends on the difference $\wde-\frac{a Q}{3 \cH \rde}$. Thus, the interaction manifests itself by changing the scale factor dependence of the dark matter as well as dark energy. 
There are different forms of the choice of the phenomenological interaction term $Q$, the models with $Q$ proportional to either $\rdc$ or $\rde$ or any combination of them are among the more popular choices,~\cite{bohmer2008prd,clemson2012prd,acosta2014prd,yang2014prd1,yang2018prd} to mention a few. It must be mentioned here that there is no particular theoretical compulsion for any of these choices. We have taken the covariant form of the source term such that it is proportional to the dark energy density ($Q^{\mu} \propto \rde$) and is parallel to the matter 4-velocity ($ Q^{\mu} \parallel u^{\mu}_{c}$) and is written as
\begin{equation} \label{eq:inter}
Q^{\mu} =   \frac{\cH \rde \, u^{\mu}_{c} \, \bela}{a}\,.
\end{equation}
Here, $\bela$ is the coupling parameter evolving with the scale factor, $a$.  The coupling parameter determines the strength of interaction and direction of energy flow; $\beta = 0 $ indicates that there is no coupling in the dark sector. In this work, we considered two possible scenarios,
\begin{description}
\item[Model L\,] If the coupling was not significant in the early Universe ($a=0$) and is felt only at the recent epoch. 
\item[Model E\,] If the interaction is predominantly an early phenomenon and is insignificant now ($a=1$). 
\end{description}
We compared the models with the Universe with a constant interaction parameter ({\bf Model C}). The ansatz chosen for the models are simple analytic functions of $a$ which are well-behaved in the region $a \in \left[0,1\right]$.
\begin{subequations}
\begin{eqnarray}
\mbox{\hypertarget{MI}{\bf Model L}\,:} \hspace{1cm} \bela &=& \beta_{0}\paren*{\frac{2\,a}{1+a}}, \label{eq:b1}\\
\mbox{\hypertarget{MII}{\bf Model E}\,:} \hspace{1cm} \bela &=& \beta_{0}\paren*{\frac{1-a}{1+a}},\label{eq:b2}\\
\mbox{\hypertarget{MIII}{\bf Model C}\,:} \hspace{1cm} \bela &=& \beta_{0}. \label{eq:b3}
\end{eqnarray}
\end{subequations}
The terms in parenthesis in the Eqns.\ (\ref{eq:b1}) and (\ref{eq:b2}) are positive definite for the domain of $a$ under consideration and hence the direction of energy flow is determined by the signature of the constant $\beta_{0}$.

It is considered in this work that the DE has a dynamical EoS parameter given by the well-known Chevallier-Polarski-Linder (CPL) parametrisation~\cite{chevallier2001ijmpd, linder2003prl} as
\begin{equation}
\wde = w_0 + w_{1} \paren*{1-a} \, ,\label{eq:w-cpl}
\end{equation}
where $w_{0}$ and $w_{1}$ are constants. A dimensionless interaction term is defined as $\Omega_{I} = \frac{Q}{3 H^{3}/\kappa}$ and the dimensionless density parameter of matter (baryonic matter  and cold dark matter (DM), denoted as `$m\paren*{=b+c}$') and dark energy (DE) are defined as $\Omega_{m}=\frac{\rdm}{3\,H^2 /\kappa}$ and $\Omega_{de}=\frac{\rde}{3\,H^2 /\kappa}$ respectively. Similarly, energy density parameter for radiation (denoted as `$r \paren*{=\gamma+\nu}$') is $\Omega_{r}=\frac{\rho_{r}}{3\,H^2 /\kappa}$. Here $H$ is the Hubble parameter defined with respect to the cosmic time $t$ and the dimensionless Hubble parameter at the present epoch is defined as $h = \frac{H_0}{100 \hskip1ex \footnotesize{\mbox{km s}^{-1} \mbox{Mpc}^{-1}}}$. The parameter values used in this work are listed in table \ref{tab:backval}, where the values are taken from the latest 2018 data release of the \Planck collaboration~\cite{planck2018cp} (\Planck 2018, henceforth).
\begin{table}[!h]
\begin{center}
\caption{\label{tab:backval}
Values of parameters used in this work based on \Planck 2018.}
\begin{tabular}{cc}
\hline \hline
Parameter&  \hspace{24ex}  Value\\
\hline
\rule[-1ex]{0pt}{2.5ex}$\Omega_b h^2$ & \hspace{24ex} $0.0223828$ \\
\rule[-1ex]{0pt}{2.5ex}$\Omega_c h^2$ & \hspace{24ex} $0.1201075$  \\
\rule[-1ex]{0pt}{2.5ex}$H_{0} \left[ \mbox{km s}^{-1} \mpci \right]$ & \hspace{24ex}   $67.32117$  \\
\hline
\hline
\end{tabular}
\end{center}
\end{table}

As shown by Pav{\'o}n and Wang~\cite{pavon2009grg}, energy transfer from dark energy to dark matter (DE $\rightarrow$ DM) is thermodynamically favoured following the Le Ch{\^a}telier-Braun principle. Observational data, on the other hand, prefer energy transfer from dark matter to dark energy (DM $\rightarrow$ DE)~\cite{zhang2012jcap, yang2018mnras, yang2018prd, yang2018jcap}. It must be noted that though the parameters $\beta_{0}$ and $\wde$ are in principle independent, they largely affect the perturbation evolutions and hence are correlated in parameter space of perturbation constraints. It had been shown in~\cite{valiviita2008jcap, he2009plb, majerotto2010mnras} that gravitational instabilities arise for constant $\wde\simeq -1$ due the interaction term in non-adiabatic pressure perturbations of dark energy.  The early time instabilities in the evolution of dark energy perturbation~\cite{valiviita2008jcap, he2009plb, gavela2009jcap, jackson2009prd, caldera2009prd, chongchitnan2009prd, xia2009prd, gavela2010jcap, clemson2012prd, mehrabi2015mnras} depend on the parameters $\beta_{0}$ and $\paren*{1+\wde}$ via a ratio called the doom factor, given as
\begin{equation}
d \equiv -\frac{aQ}{3\cH\rde\paren*{1+\wde}}.
\end{equation}
To avoid early time instabilities, $d$ must be negative semi-definite ($d \le 0$)~\cite{gavela2009jcap}, ensuring that $\beta_{0}$ and $\paren*{1+\wde}$ have the same sign. Thus stable perturbations can be achieved with either energy flow from dark matter to dark energy ($\beta_{0}>0$) and non-phantom or quintessence EoS ($\paren*{1+\wde}>0$) or energy flow from dark energy to dark matter ($\beta_{0}<0$) and phantom EoS ($\paren*{1+\wde}<0$). 

\begin{figure*}[!htbp]
        \centering
            \subfloat{\includegraphics[width=.5\linewidth]{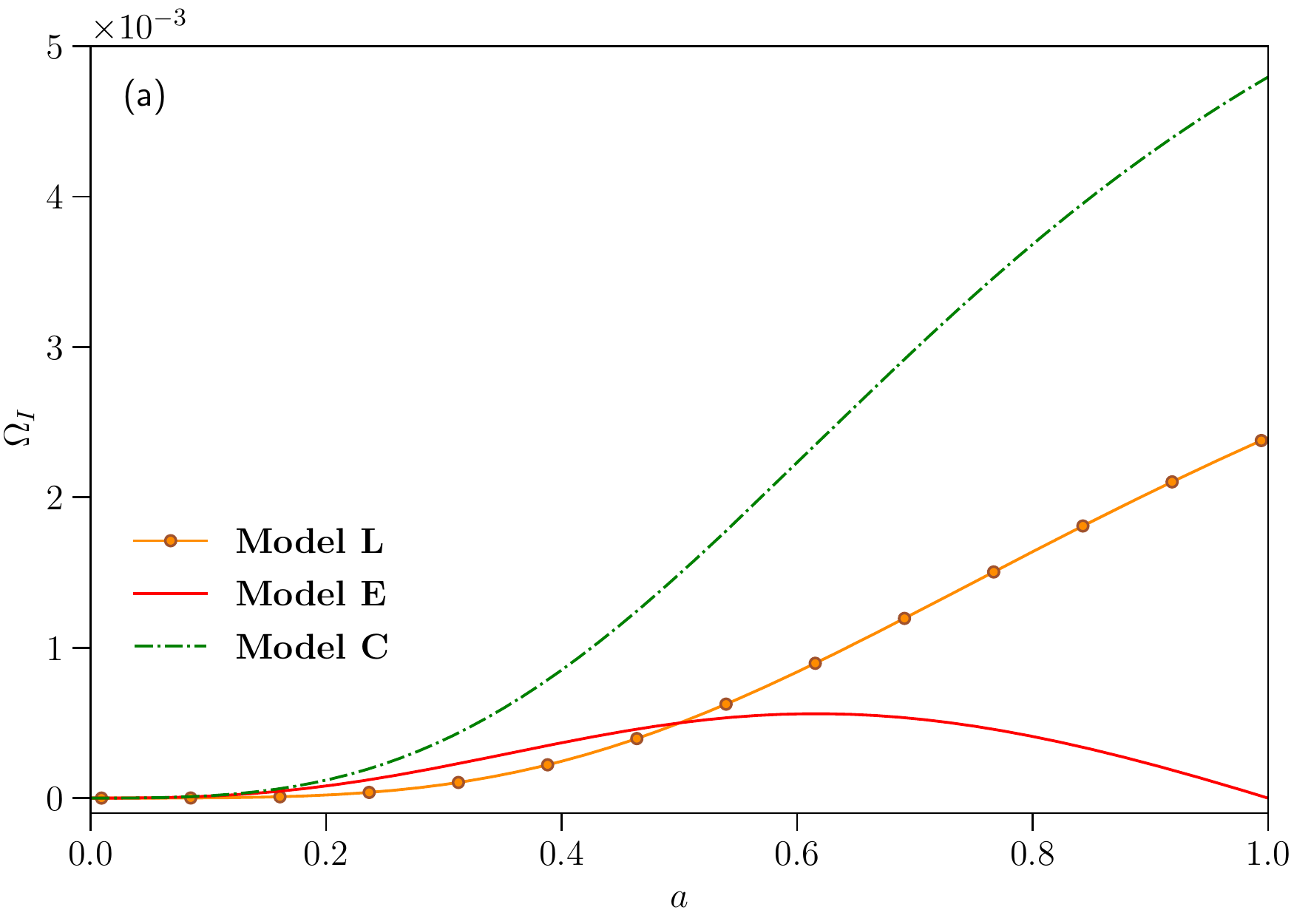}}\hfill
            \subfloat{\includegraphics[width=.5\linewidth]{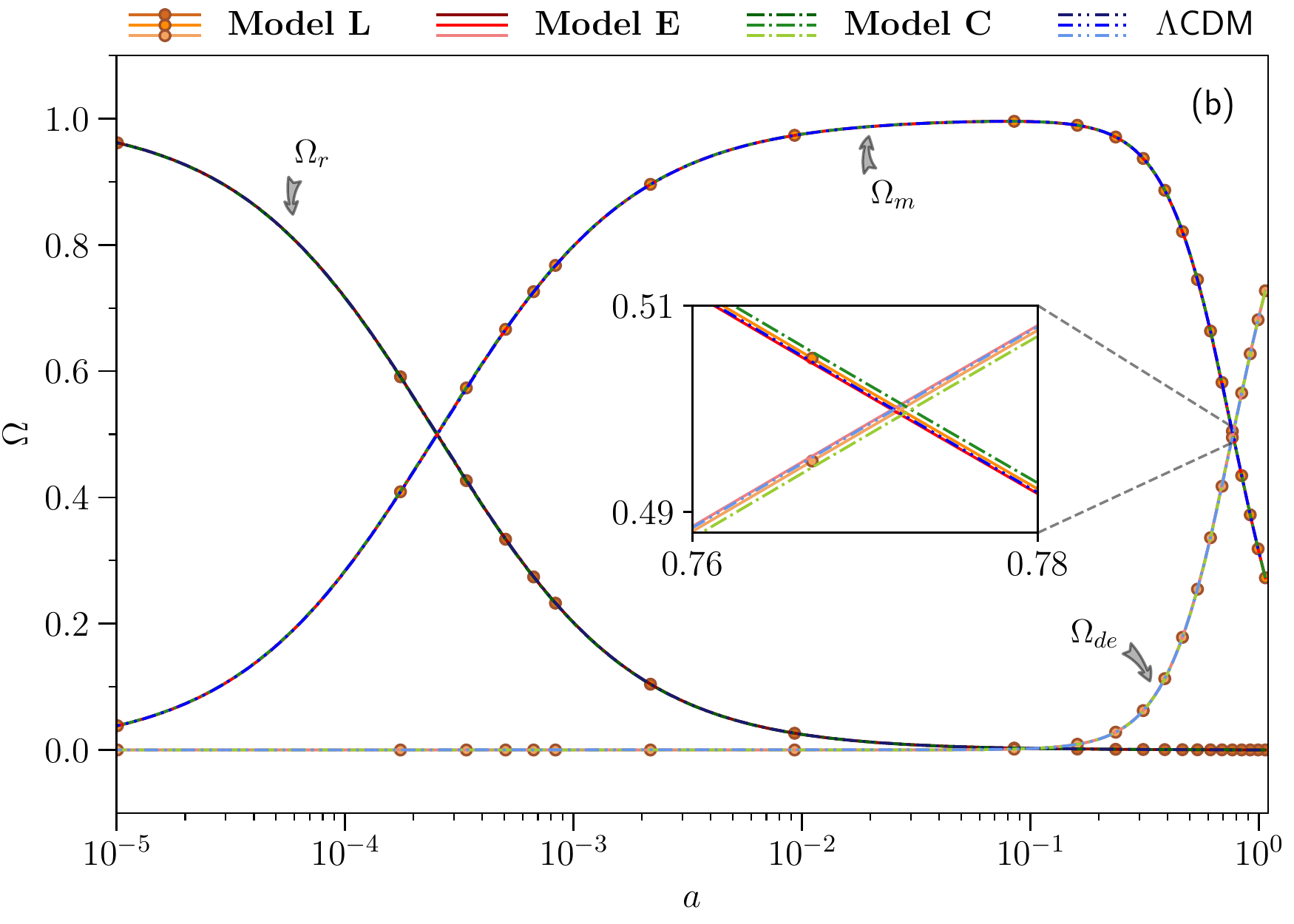}}\hfill
        \caption{Plot of (a) the dimensionless interaction parameter $\Omega_{I}$ and (b) density parameter $\Omega$ against scale factor $a$. The x-axis in Fig.\ (b) is in logarithmic scale The solid line with solid circles represents {\bf Model L}, solid line represents {\bf Model E} and dashed-dot line represents {\bf Model C} while the dashed-dot-dot line is for \lcdm. The inset shows the zoomed-in portion for the region $a= 0.76$ to $a = 0.78$.}\label{im:bck2}
\end{figure*}

In this section and the next (Sect.\ \ref{sec:pert}), we have considered the energy flow from dark matter to dark energy and $\beta_{0}$ to be positive and hence $\wde>-1$. We have chosen the magnitude of $\beta_{0}$ to be small consistent with the observational results given in~\cite{yang2014prd1, yang2017prd1, yang2017prd2, pan2018mnras, vagnozzi2020mnras}. The particular value used here, $\beta_{0}=0.007$, is an example chosen such that no instability in the dark energy perturbation arises. For the background and perturbation analyses (Sect.\ \ref{sec:pert}), we have chosen the example values of the parameter, $w_{0}$ and $w_{1}$ in $\wde$ (Eqn.\ (\ref{eq:w-cpl})) as
\begin{equation}\label{eq:w}
w_{0} = -0.9995, ~~w_{1} = 0.005.
\end{equation}
The chosen values of the parameters $w_{0}$ and $w_{1}$ also ensure that $w_{de} \sim -1$ at $a=1$. It must be mentioned that, EoS parameter in the quintessence region is considered solely to avoid DE models with a  future ``big-rip'' singularity associated with phantom EoS parameter. Several instances of interacting DE models with $\wde <-1$ are found in the literature~\cite{valiviita2008jcap, gavela2009jcap, divalentino2017prd1, divalentino2017prd2, yang2017prd1, yang2017prd2, pan2018mnras, yang2018prd, yang2018jcap}. Figure (\ref{im:bck2}a) shows the evolution of $\Omega_{I}$ with scale factor $a$ for \modellate, \modelearly and \modelcons. In Fig.\ (\ref{im:bck2}a) the direction of energy flow is from dark matter to dark energy and the magnitude of $\Omega_{I}$ is the rate of energy transfer. The variation of density parameters of radiation ($\Omega_{r}$), dark matter together with baryons ($\Omega_{m}$) and dark energy ($\Omega_{de}$) with scale factor $a$ in logarithmic scale is shown in Fig.\ (\ref{im:bck2}b) for the three models and the \lcdm model. It is clear from Figs.\ (\ref{im:bck2}a) and (\ref{im:bck2}b) that the effect of interaction will be very small in its contribution to the density parameters, $\Omega_{A}$, where $A = r,\,m,\,de$.

\section{Evolution of perturbations}\label{sec:pert}
The perturbed FLRW metric in a general gauge takes the form~\cite{kodama1984ptps, mukhanov1992pr, ma1995apj} 
\begin{equation} \label{eq:metric2}
\begin{split}
ds^2=a^2\paren*{\tau} & \left\{ -\paren*{1+2\phi}d\tau^2+2\,\partial_iB \,d\tau \,dx^i +\right. \\
& \left.\left[\paren*{1-2\psi}\delta_{ij}+2\partial_i\partial_jE\right]dx^idx^j \right\},
\end{split}
\end{equation}
where $\phi, \psi, B, E$ are gauge-dependant scalar functions of space and time. In presence of interaction, the covariant form of the energy-momentum conservation equation will be
\begin{equation} \label{eq:condition}
T^{\,\mu \nu}_{\paren*{A}; \nu} =Q^{\,\mu}_{\paren*{A}} \,, \hspace{0.3cm} \mbox{where}  \hspace{0.2cm} \sum_{A}Q^{\mu}_{\paren*{A}} =0 ~.
\end{equation}
The energy-momentum transfer function for the fluid `$A$', $Q^{\,\mu}_{\paren*{A}}$, can be split into the energy transfer rate, $Q_{\paren*{A}}$ and the momentum transfer rate, $F^{\mu}_{\paren*{A}}$, relative to the total $4$-velocity as~\cite{valiviita2008jcap, majerotto2010mnras, clemson2012prd}
\begin{equation} \label{Q-pert-def}
Q^{\,\mu}_{\paren*{A}} = Q_{\paren*{A}} u^{\,\mu} + F^{\,\mu}_{\paren*{A}}\,, \hspace{0.5cm} u_{\mu} F^{\,\mu}_{\paren*{A}} =0\,, \hspace{0.5cm} F^{\,\mu}_{\paren*{A}}= a^{-1}\, \paren*{ 0,\partial^{i}\, f_A}.
\end{equation}
Writing the total 4-velocity, $u^{\,\mu}$, in terms of the total peculiar velocity, $v$ as 
\begin{equation}\label{eq:v4}
u^{\,\mu} = a^{-1}\paren*{1-\phi, v^{i}},
\end{equation}
the temporal and spatial components of the 4-energy-momentum transfer rate can be written as
\begin{eqnarray}
Q^{0}_{\paren*{A}}&=&a^{-1}\,\left[Q_A(1-\phi)+\delta Q_A\right], \\
\hspace{0.5cm} \mbox{and} \hspace{0.5cm} Q^{i}_{\paren*{A}} &=&a^{-1}\,\left[Q_A\,v^{i}+ \partial^{i}\,f_A\right]
\label{eq:Q-component}
\end{eqnarray}
respectively, where $\delta Q_A$ is the perturbation in the energy transfer rate and $f_A$ is the momentum transfer potential.

The perturbed conservation equations of the fluid `$A$' in the Fourier space are written as
\begin{eqnarray}
\begin{split}
\delta \rho^\prime_A  -3\paren*{\rA+\pA}\psi'+ &k\paren*{\rA+\pA}\paren*{\vA+E^{\prime}} + \\3 \cH \,\paren*{\delta \rA+\delta \pA} =&~aQ_A\phi +a \delta Q_A , \label{eq:e1}
\end{split}\\
\begin{split}
\left[\paren*{\rA+\pA}\paren*{\vA+B}\right]'+ &4\cH \paren*{\rA+\pA}\paren*{\vA+B}-\\
k\paren*{\rA+\pA}\phi-k\,\delta \pA &=aQ_A\paren*{v+B}-a\, kf_A ~. \label{eq:m1}
\end{split}
\end{eqnarray}\\
In Eqns.\ (\ref{eq:e1}) and (\ref{eq:m1}), $\delta \rA$ is the perturbation in the energy density, $\delta \pA$ is the perturbation in pressure, $u^{\,\mu}_{A} = a^{-1}\paren*{1-\phi, \vA^{i}}$ is the 4-velocity with peculiar velocity $\vA$ of the fluid `$A$' and $k$ is the wavenumber. For an adiabatic perturbation, the pressure perturbation~\cite{wands2000prd, malik2003prd, malik2005jcap, valiviita2008jcap, malik2009pr} in presence of interaction is
\begin{equation}\label{eq:pert-p}
\delta \pA=c_{s,\,A}^2 \delta \rA+\paren*{c_{s,\,A}^2-c_{a,\,A}^2}\left[3 \cH\paren*{1+w_A}\rA -a Q_A\right]\frac{\vA}{k},
\end{equation}
where $c_{a,\,A}^2=\frac{\pA^\prime}{\rA^\prime}$ is the square of adiabatic sound speed and $c_{s,\,A}^2=\frac{\delta \pA}{\delta \rA}$ is the square of effective sound speed in the rest frame of $A$-th fluid. 

The dynamical coupling parameter $\beta_{0}$ defined in Eqn.\ (\ref{eq:inter}) in the previous section is considered to be not affected by  perturbation. This assumption is valid for the EoS parameter defined in Eqn.\ (\ref{eq:w-cpl}) and the Hubble parameter, $\cH$. These perturbation equations are solved along with the perturbation equations~\cite{kodama1984ptps, mukhanov1992pr, ma1995apj} of the radiation, neutrino and baryon using the publicly available Boltzmann code \camb\footnote{Available at: \href{https://camb.info}{https://camb.info}}~\cite{lewis1999bs} after suitably modifying it. 

Using (\ref{eq:v4}), Eqn.\ (\ref{eq:inter}) can be conveniently written as
\begin{equation} \label{eq:q1}
Q= \frac{\cH \rde\, \beta\paren*{a}}{a}.
\end{equation}
Defining the density contrasts of the dark matter and dark energy as $\ddc = \delta\rdc/\rdc$ and $\dde = \delta\rde/\rde$ respectively and using Eqns.\ (\ref{eq:pert-p}) and (\ref{eq:q1}), the perturbation Eqns.\ (\ref{eq:e1}) and (\ref{eq:m1}) are written in synchronous gauge~\cite{ma1995apj} ($\phi=B=0$, $\psi=\eta$ and $k^2\,E=-\msh/2-3\eta$, where $\eta$ and $\msh$ are synchronous gauge fields in the Fourier space) as
\begin{widetext}
\begin{eqnarray}
\ddc^\prime+ k v_{c} +\frac{\msh^\prime}{2} &=& \cH \bela \frac{\rde}{\rdc}\paren*{\ddc-\dde}, \\ \label{eq:e2dm}
v_{c}^\prime+\cH v_{c}&=& 0~, \label{eq:m2dm}
\end{eqnarray}
\begin{equation}
\begin{split}
\dde^\prime +3  \cH & \paren*{\cde-\wde}\dde+\paren*{1+\wde}\paren*{k \vde+\frac{\msh^\prime}{2}}\\
+3 \cH & \left[3 \cH \paren*{1+\wde}\paren*{\cde-\wde}\right]\frac{\vde}{k} +3\cH \wde^\prime\frac{\vde}{k} \\
=\, 3 \cH^{2} & \bela  \paren*{\cde-\wde}\frac{\vde}{k}, \label{eq:e2de}
\end{split}
\end{equation}
\begin{equation}
\vde^\prime+\cH\paren*{1-3\cde}\vde-\frac{k\,\dde\,\cde }{\paren*{1+\wde}}=\frac{\cH\, \bela}{\paren*{1+\wde}}\, \left[\vdc-\paren*{1+\cde}\vde\right]. \label{eq:m2de}
\end{equation}
\end{widetext}
It may be noted that although the interaction term in Eqn.\ \ref{eq:q1} is similar to that used in~\cite{kumar2019epjc}, the perturbation equations (Eqns.\ (\ref{eq:e2dm})-(\ref{eq:m2de})) are different from those in~\cite{kumar2019epjc} as we have not considered vacuum energy with $\wde = -1$. For the same reason the initial conditions, to follow, are also different in our case. For a detailed discussion on perturbation equations in an inhomogeneous vacuum scenario, we refer to~\cite{wands2012cqg, desantiago2012ax, wang2013prd, salvatelli2014prl, martinelli2019mnras}. The coupled differential equations (Eqns.\ (\ref{eq:e2dm})-(\ref{eq:m2de})) are solved with $k = 0.1\,h$ $\mpci$ and the adiabatic initial conditions using \camb. Using the gauge-invariant quantity~\cite{malik2003prd, malik2009pr, he2009plb, chongchitnan2009prd, xia2009prd} $\zeta_{A} = \paren*{-\psi - \cH \frac{\delta \rho_{A}}{\rho_{A}^{\prime}}}$ and relative entropy perturbation $S_{AB} = 3\paren*{\zeta_{A}-\zeta_{B}}$, the adiabatic initial conditions for $\ddc$, $\dde$ in presence of interaction are obtained respectively as 
\begin{subequations}
\begin{eqnarray}
\delta_{ci} &=& \left[3 +\frac{\rde}{\rdc}\bela\right] \frac{\delta_{\gamma}}{3\paren*{1+w_{\gamma}}}, \label{eq:initial-m}\\
\delta_{dei} &=& \left[3\,\paren*{1+\wde} - \bela \right] \frac{\delta_{\gamma}}{3\paren*{1+w_{\gamma}}}, \label{eq:initial-de}
\end{eqnarray}
\end{subequations}
Here, $\delta_{\gamma}$ is the density fluctuation of photons. As can be seen from Eqn.\ (\ref{eq:m2dm}), there is no momentum transfer in the DM frame, hence initial value for $v_{c}$ is set to zero ($v_{ci} =0$)~\cite{bean2008prd2, chongchitnan2009prd, xia2009prd, valiviita2008jcap}. The initial value for the dark energy velocity, $\vde$ is assumed to be same as the initial photon velocity, $v_{dei} = v_{\gamma\,i}$.
To avoid the instability in dark energy perturbations due to the the propagation speed of pressure perturbations, we have set $\cde = 1$~\cite{waynehu1998apj, bean2004prd, gordon2004prd, afshordi2005prd, valiviita2008jcap}.

\begin{figure*}[!htbp]
  \centering
\includegraphics[width=\textwidth]{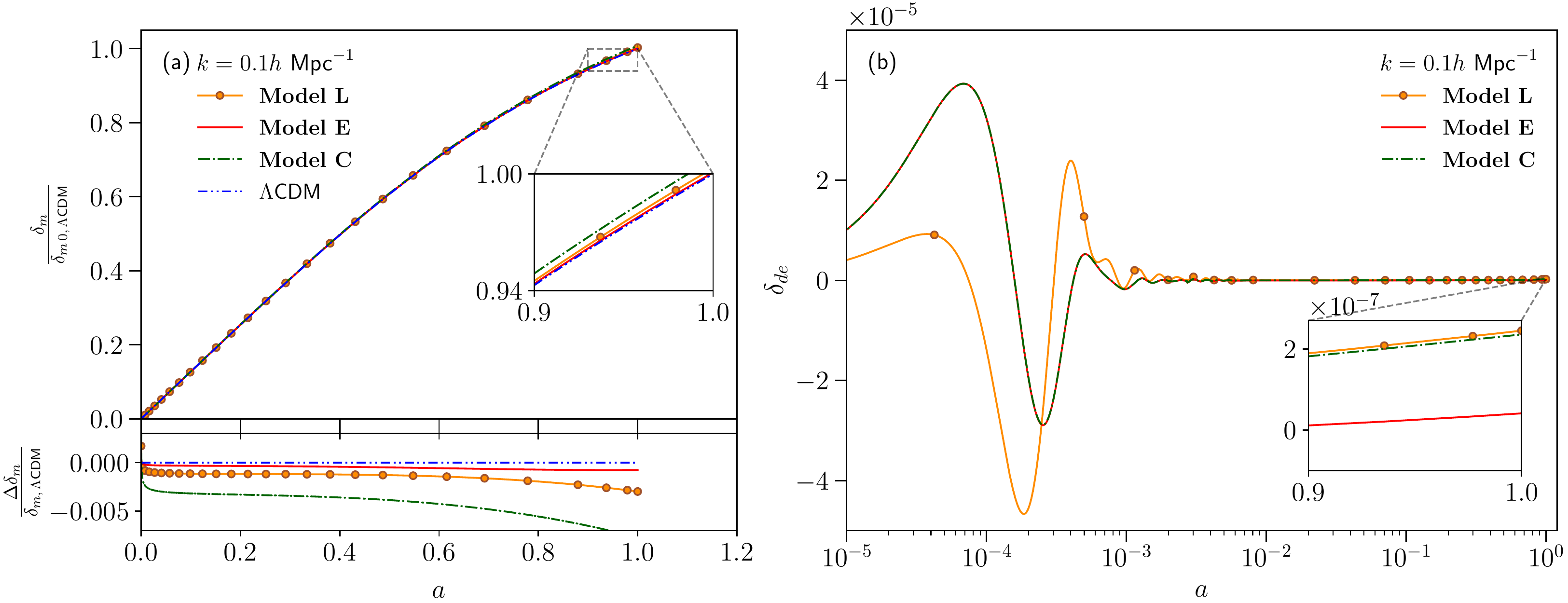}
\caption{(a) Plot of \textbf{Upper Panel} : the matter density contrast $\frac{\delta_{m}}{\delta_{m\,0,\,\scriptsize{\Lambda\text{CDM}}}}$ and \textbf{Lower Panel} : fractional growth rate is defined as $\frac{\Delta \delta_{m}}{\delta_{m,\,\scriptsize{\Lambda\text{CDM}}}} = \paren*{1- \frac{\delta_{m}}{\delta_{m,\,\scriptsize{\Lambda\text{CDM}}}}}$ relative to the \lcdm model against $a$. The origin on the x-axis represents $10^{-5}$. (b) Plot of the dark energy density fluctuation, $\dde$ against $a$ in logarithmic scale for $k = 0.1\,h$ $\mpci$. The solid line with solid circles represents {\bf Model L}, solid line represents {\bf Model E} and dashed-dot line represents {\bf Model C} while the dashed-dot-dot line is for \lcdm. The inset shows the zoomed-in portion from $a = 0.9$ to $a = 1.0$.}\label{im:delta1}
\end{figure*}

Figure (\ref{im:delta1}a) shows the variation of the density contrast, $\ddm = \delta \rdm/\rdm$ for the cold dark matter ($c$) taken together and the baryonic matter ($b$) against $a$ for \modellate, \modelearly and \modelcons along with the \lcdm model. For a better comparison with the \lcdm model, $\ddm$ is scaled by $\delta_{m0} = \ddm\paren*{a=1}$ of \lcdm\footnote{The origin on the x-axis is actually $10^{-5}$}. As can be seen from the Fig.\ (\ref{im:delta1}a), the growth of density fluctuation $\ddm$ is similar in all the model at early times. The effect of interaction comes into play at late time. The late-time growth of $\ddm$ (inset of (\ref{im:delta1}a)) shows that \modelearly agrees well with the \lcdm model, whereas \modellate and \modelcons grow to a little higher value. Figure (\ref{im:delta1}b) shows the variation of the dark energy density contrast $\dde$ for \modellate, \modelearly and \modelcons. At early time, $\dde$ oscillates and then decays to very small values. In \modelcons, the early time evolution of $\dde$ is similar to \modelearly while the late time evolution is similar to \modellate. To understand the differences among the three models and the \lcdm model, we have shown the fractional matter density contrast, $\frac{\Delta \delta_{m}}{\delta_{m,\,\scriptsize{\Lambda\text{CDM}}}} = \paren*{1- \frac{\delta_{m}}{\delta_{m,\,\scriptsize{\Lambda\text{CDM}}}}}$ in the lower panel of Fig.\ (\ref{im:delta1}a). It is clearly seen that, $\delta_{m}$ for \modelearly evolves close to the \lcdm model. 

\subsection{Effect on CMB temperature, matter power spectrum and $\fsg$}\label{sec:result}
It is necessary to have an insight into other physical quantities like the CMB temperature spectrum, matter power spectrum and the logarithmic growth of matter perturbation, to differentiate the interacting models. The CMB temperature power spectrum is given as
\begin{equation}
C_{\ell}^{TT} = \frac{2}{k} \int k^{2} d k \,P_{\zeta}\paren*{k} \Delta^{2}_{T\ell}\paren*{k},
\end{equation}
where $\ell$ is the multipole index, $P_{\zeta}\paren*{k}$ is the primordial power spectrum, $\Delta_{T\ell}\paren*{k}$ is the temperature transfer function and $T$ represents the temperature. For a detailed analysis on the CMB spectrum we refer to~\cite{hu1995apj,seljak1996apj,dodelson2003}. The matter power spectrum is written as
\begin{equation} \label{eq:power}
P\left(k,a\right)= A_s \,k^{n_s} T^2\left(k\right) D^2\left(a\right),
\end{equation}
where $A_{s}$ is the scalar primordial power spectrum amplitude, $n_{s}$ is the spectral index, $T\left(k\right)$ is the matter transfer function and $D\left(a\right)=\frac{\ddm\left(a\right)}{\ddm\left(a=1\right)}$ is the normalised density contrast. For a detailed description we refer to~\cite{dodelson2003}. Both $C_{\ell}^{TT}$ and $P\paren*{k,a}$ are computed numerically using \camb. The values of power spectrum amplitude, $A_{s} = 2.100549 \times 10^{-9}$ and spectral index, $n_{s} = 0.9660499$ are taken from \Planck 2018 data~\cite{planck2018cp}.

\begin{figure*}[!htbp]
\centering
\includegraphics[width=\textwidth]{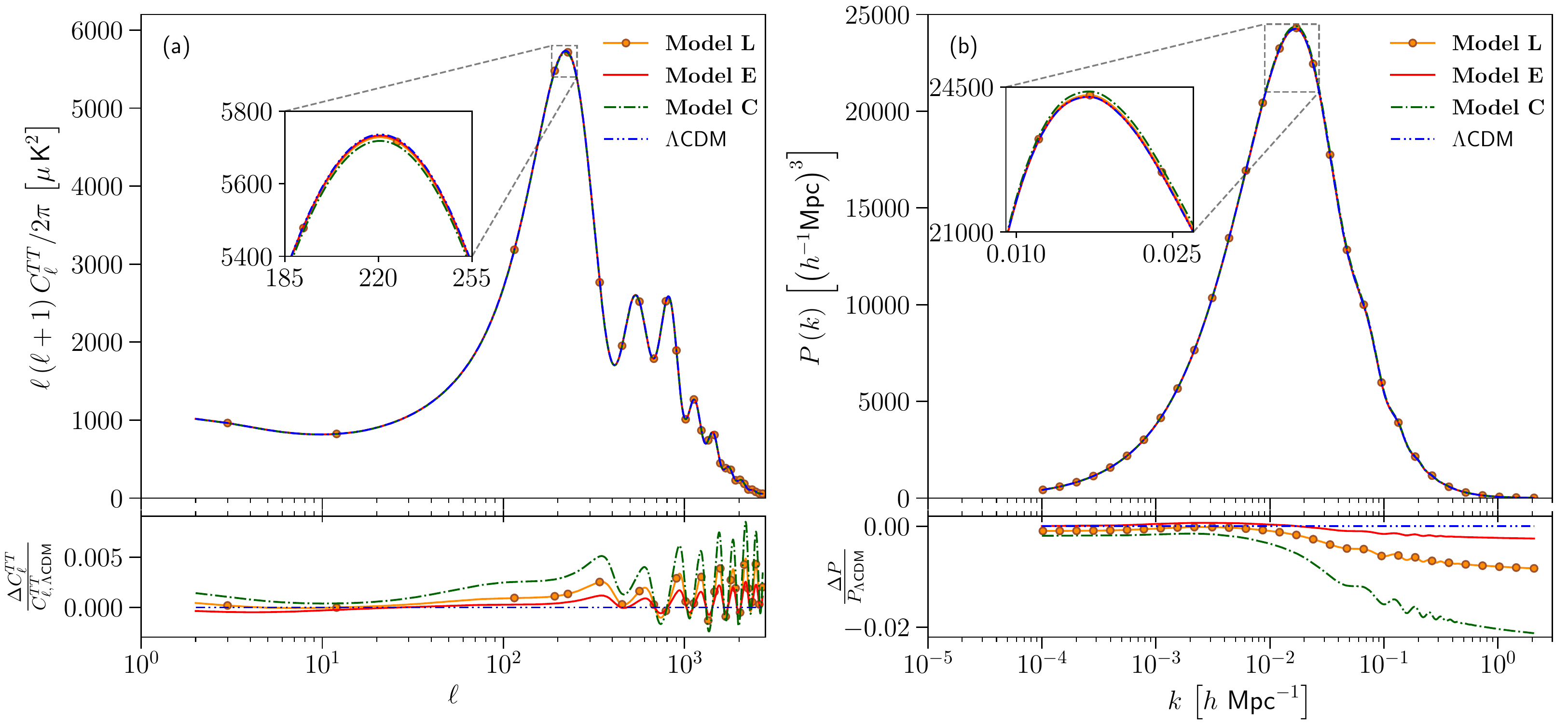}
\caption{\textbf{Upper Panel} : (a) Plot of CMB temperature power spectrum in units of $\mu \mbox{K}^{2}$ with the multipole index $\ell$ in logarithmic scale. (b) Plot of matter power spectrum $P\paren*{k}$ in units of $\paren*{h^{-1}\mpc}^{3}$ with wavenumber $k$ in units of $h\,\mpci$. \textbf{Lower Panel} : Plot of fractional change in the temperature spectrum, $\frac{\Delta C_{\ell}^{TT}}{C_{\ell,\, \scriptsize{\Lambda\text{CDM}}}^{TT}} = \paren*{1- \frac{C_{\ell}^{TT}}{C_{\ell,\, \scriptsize{\Lambda\text{CDM}}}^{TT}}}$ and the fractional change in matter power spectrum, $\frac{\Delta P}{P_{\scriptsize{\Lambda\text{CDM}}}} = \paren*{1- \frac{P}{P_{\scriptsize{\Lambda\text{CDM}}}}}$. For both the panels, the solid line with solid circles represents {\bf Model L}, solid line represents {\bf Model E} and dashed-dot line represents {\bf Model C} while the dashed-dot-dot line is for \lcdm at $a=1$. The inset shows the zoomed-in versions of the peaks.}\label{im:cl}
\end{figure*}
Figure (\ref{im:cl}) shows the temperature and matter power spectrum for \modellate, \modelearly, \modelcons and \lcdm at $a=1$. In \modellate and \modelcons, more matter content results in lower amplitude of the first peak of the CMB spectrum compared to the \lcdm model. The lower panel of Fig.\ (\ref{im:cl}a), shows the fractional change ($=\Delta C_{\ell}^{TT}/C_{\ell,\, \scriptsize{\Lambda\text{CDM}}}^{TT}$) in $C_{\ell}^{TT}$. It is seen from the lower panel of Fig.\ (\ref{im:cl}a), that the low-$\ell$ modes of \modelearly increases through the integrated Sachs-Wolfe (ISW) effect. More matter content also increases the matter power spectrum compared to the \lcdm model. The deviations from the \lcdm model are prominent for the smaller modes. These features are clear from the lower panel of Fig.\ (\ref{im:cl}b), which shows the  fractional change in matter power spectrum, $\Delta P/P_{\scriptsize{\Lambda\text{CDM}}}$ of the interacting models relative to the \lcdm model.

The presence of the interaction modifies the logarithmic growth rate which helps in differentiating between the models even better. The growth rate is the logarithmic derivative of the density fluctuation of matter (baryon and CDM) and is written as
\begin{equation}
f\paren*{a}= \frac{d \ln \ddm}{d \ln a} ~=~ a \frac{d}{d\,a}\paren*{\frac{\delta \rdm}{\rdm}}.
\end{equation}
Since, $\delta \rdm = \paren*{\delta_{c}\rdc + \delta_{b}\rho_{b}}$, $\delta_{b}$ being the baryon density fluctuation, in presence of interaction the growth rate~\cite{costa2017jcap} will be
\begin{equation} \label{eq:growth_rate}
f\paren*{a} = a \paren*{\frac{\delta_{c,\,a} \,\, \rdc + \delta_{b,\,a} \,\,\rho_{b}}{\ddm \, \rdm} - \frac{a Q\, \ddc}{\ddm \,\rdm} - \frac{a Q}{\rdm}},
\end{equation}
where `$_{,\,a}$' denotes the derivative with respect to the scale factor $a$ and $Q$ is given by Eqn.\ (\ref{eq:q1}). It must be noted that the last two terms involving interaction $Q$ is introduced in Eqn.\ (\ref{eq:growth_rate}) via the evolution of $\rdc$ (Eqns.\ (\ref{eq:con1})). We have calculated the growth rate, $f$ for the different models using \camb.

Observationally the galaxy density fluctuation, $\delta_{g}$ is measured, which in turn gives the matter density fluctuation, $\ddm$ as $\delta_{g} = b \ddm$, where $b \in \left[1,3\right]$ is the bias parameter. This $\ddm$ is used to calculate the logarithmic growth rate, $f$. Thus, $f$ is sensitive to $b$ and is not a very reliable quantity. A more dependable observational quantity is defined as the product $f\paren*{a}\se\paren*{a}$~\cite{percival2009mnras}, where $\se\paren*{a}$ is the root-mean-square (rms) mass fluctuations within the sphere of radius $R = 8 h^{-1}\, \mpc$. The mean-square mass fluctuation is given by
\begin{equation} \label{eq:sigma2}
\sigma^2\left(R,z\right)=\frac{1}{2 \pi^2} \int k^3 P\left(k,z\right) W\left( kR\right)^2 \frac{d k}{k}
\end{equation}
where $P\left(k,z\right)$ is the power spectrum given in Eqn.\ (\ref{eq:power}) and $W\left( kR\right)$ is the top-hat window function given by
\begin{equation} \label{eq:window}
W\left( kR\right)=3\left[\frac{\sin\left(k R\right)}{\left(k R\right)^3}-\frac{\cos\left(k R\right)}{\left(k R\right)^2}\right].
\end{equation}
When the size of the filter is $R = 8 h^{-1}\, \mpc$, $\sigma^{2}\paren*{R,z} \equiv \sigma_{8}^{2}\paren*{z}$. The rms linear density fluctuation is also written as $\se\paren*{a} = \se\paren*{1}\frac{\ddm\paren*{a}}{\ddm\paren*{1}}$, where $\se\paren*{1}$ and $\ddm\paren*{1}$ are the values at $a=1$, and $f$ and $\se\paren*{1}$ for the different models are obtained from Eqn.\ (\ref{eq:growth_rate}) and (\ref{eq:sigma2}) using our modified version of \camb. The combination $\fsg$ is written as
\begin{equation} \label{eq:fsigma}
\fsg\paren*{a} \equiv f\paren*{a}\se\paren*{a} = \se\paren*{1}\frac{a}{\ddm\paren*{1}}\frac{d \ddm}{d \,a}.
\end{equation}

\begin{figure*}[!htbp]
\centering
\includegraphics[width=\textwidth]{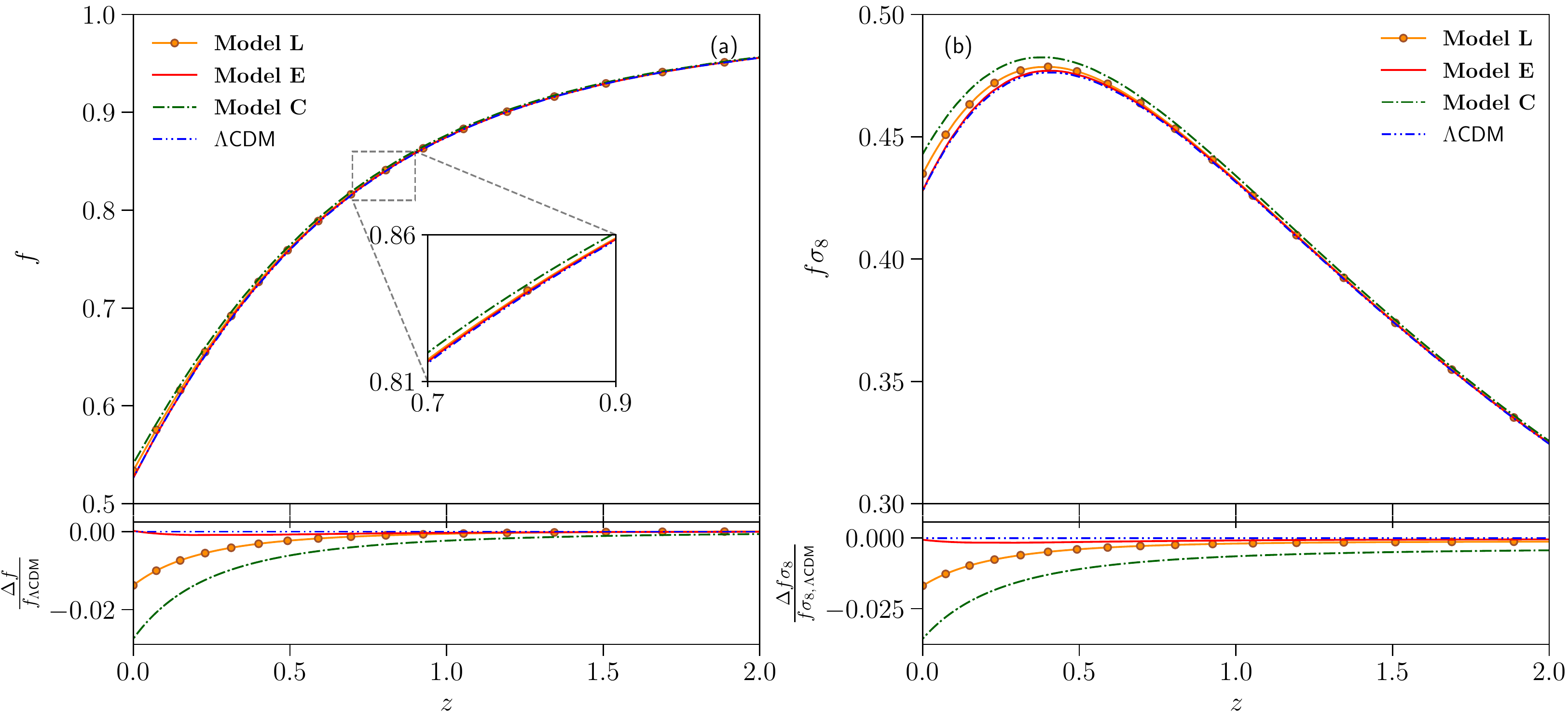}
\caption{\textbf{Upper Panel} : Plot of (a) linear growth rate $f$ and (b) $\fsg$ against redshift $z$. The inset shows the zoomed-in portion from $z = 0.7$ to $z = 0.9$. \textbf{Lower Panel} : Plot of fractional change in the temperature spectrum, $\frac{\Delta f}{f_{\scriptsize{\Lambda\text{CDM}}}} = \paren*{1- \frac{f}{f_{\scriptsize{\Lambda\text{CDM}}}}}$ and the fractional change in matter power spectrum, $\frac{\Delta \fsg}{f\sigma_{8,\,\scriptsize{\Lambda\text{CDM}}}} = \paren*{1- \frac{\fsg}{f\sigma_{8,\,\scriptsize{\Lambda\text{CDM}}}}}$. For both the panels, the solid line with solid circles represents {\bf Model L}, solid line represents {\bf Model E} and dashed-dot line represents {\bf Model C} while the dashed-dot-dot line is for \lcdm.}\label{im:fs}
\end{figure*}
The logarithmic growth rates, $f$ and $\fsg$ are independent of the wavenumber $k$ for smaller redshift, $z$, so only the domain $z=0$ to $z=2$ is considered here. The redshift, $z$ and the scale factor $a$ are related as $z = \frac{a_0}{a} -1$, $a_0$ being the present value (taken to be unity). The difference in the models is magnified in the $f$ and $\fsg$ analysis. As can be seen from Fig.\ (\ref{im:fs}), growth rates ($f$) and $\fsg$ are different for the different models in the recent past. The differences due to the evolution of interaction are seen in $f$ and $\fsg$. Both \modellate and \modelcons have slightly higher values of $f$ and $\fsg$ at $z=0$, compared to the value obtained from the \lcdm model. For \modelearly and the \lcdm model, the values of $f$ and $\fsg$ are same at $z=0$. When the energy transfer rates were different in the recent past, \modelearly had a slightly larger value of $f$ and $\fsg$ (compared to the \lcdm model) when the interaction was non-zero. The fractional changes in growth rate ($\Delta f/f_{\scriptsize{\Lambda\text{CDM}}}$) and $\fsg$ ($\Delta \fsg/f\sigma_{8,\,\scriptsize{\Lambda\text{CDM}}}$) of the interacting models relative to the \lcdm model are shown in the lower panels. The difference among the three models is distinctly seen.

\section{Observational Constraints}\label{sec:mcmc}
In this section, \modellate, \modelearly and \modelcons are tested against observational datasets like the CMB, BAO, Supernovae and RSD data by using the Markov Chain Monte Carlo (MCMC) analysis of the publicly available, efficient MCMC simulator \cosmomc\footnote{Available at: \href{https://cosmologist.info/cosmomc/}{https://cosmologist.info/cosmomc/}}~\cite{lewis2013hha,lewis2002ah}. The datasets and the methodology are discussed in the Appendix \ref{sec:app}. The datasets are used to constrain the nine-dimensional parameter space given as 
\begin{equation}\label{eq:parameter}
P \equiv \lbrace \Omega_{b} h^2, \Omega_{c} h^2, 100\theta_{MC}, \tau, \beta_0, w_0, w_{1}, \ln\paren*{10^{10} A_s}, n_{s}\rbrace,
\end{equation}
where $\Omega_b h^2$ is the baryon density, $\Omega_c h^2$ is the cold dark matter density, $\theta_{MC}$ is the angular acoustic scale, $\tau$ is the optical depth, $\beta_{0}$, $w_{0}$ and $w_{1}$ are the free model parameters, $A_{s}$ is the scalar primordial power spectrum amplitude and $n_{s}$ is the scalar spectral index. The parameter space, $P$, for all the three models, is explored for the flat prior ranges given in Table \ref{tab:prior}. We allowed the prior of $\beta_{0}$ to cross the zero and set the prior of $w_{0}$ and $w_{1}$ such that such that $\wde$ is always in the quintessence region.
\begin{table}[!h]
\begin{center}
\caption{Prior ranges of nine independent parameters used in the \cosmomc analysis.}\label{tab:prior}
\begin{tabular}{cc}
\hline \hline
\rule[-1ex]{0pt}{2.5ex}Parameter &  \hspace{24ex} Prior\\
\hline
\rule[-1ex]{0pt}{2.5ex}$\Omega_{b} h^2$&  \hspace{24ex} $\left[0.005, 0.1\right]$ \\
\rule[-1ex]{0pt}{2.5ex}$\Omega_{c} h^2$&  \hspace{24ex} $\left[0.001, 0.99\right]$\\
\rule[-1ex]{0pt}{2.5ex}$100\theta_{MC}$&  \hspace{24ex} $\left[0.5, 10\right]$\\
\rule[-1ex]{0pt}{2.5ex}$\tau$&  \hspace{24ex} $\left[0.01, 0.8\right]$\\
\rule[-1ex]{0pt}{2.5ex}$\beta_0$&  \hspace{24ex} $\left[-1.0, 1.0\right]$\\
\rule[-1ex]{0pt}{2.5ex}$w_0$&  \hspace{24ex} $\left[-0.9999, -0.3333\right]$\\
\rule[-1ex]{0pt}{2.5ex}$w_{1}$&  \hspace{24ex} $\left[0.005, 1.0\right]$\\
\rule[-1ex]{0pt}{2.5ex}$\ln\paren*{10^{10} A_s}$&  \hspace{24ex} $\left[1.61, 3.91\right]$\\
\rule[-1ex]{0pt}{2.5ex}$n_{s}$&  \hspace{24ex} $\left[0.8, 1.2\right]$\\
\hline \hline
\end{tabular}
\end{center}
\end{table}

\subsection{Model L}
For \modellate, the marginalised values with errors at $1\sigma$ ($68\%$ confidence level) of the nine free parameters and three derived parameters, $H_{0}$, $\Omega_{m}$ and $\se$, are listed in Table \ref{tab:mean-1}. Henceforth, the 1D marginalised values given in the tables will be referred to as mean values. The correlations between the model parameters ($\beta_{0}$, $w_{0}$, $w_{1}$) and the derived parameters ($H_{0}$, $\Omega_{m}$, $\se$) and their marginalised contours are shown in Fig.\ \ref{im:tri1}. The contours contain $1\sigma$ region ($68\%$ confidence level) and $2\sigma$ region ($95\%$ confidence level). When only the \Planck data is considered, the mean value of the coupling parameter, $\beta_{0}(=0.00788^{+0.00815+0.0158}_{-0.00815-0.0162})$, is positive with zero in the $1\sigma$ region indicating energy transfers from DM to DE. The parameters $w_{0} (<-0.909< -0.800 )$ and $w_{1} (< 0.174 < 0.365)$ remain unconstrained even in the $2\sigma$ region. For other parameters, the mean values are compared with their \lcdm counterparts from the \Planck estimation~\cite{planck2018cp}. The Hubble expansion rate, $H_{0}$, is obtained at a value lower than $67.36\pm 0.54$ in $\mbox{km s}^{-1} \mbox{Mpc}^{-1}$, as obtained for the \lcdm model~\cite{planck2018cp}. Though the mean value is lower than that obtained from the \Planck estimate, the presence of high error bars results in $3.5\sigma$ tension with the local measurement as $H_{0} = 74.03\pm1.42$ $\mbox{km s}^{-1} \mbox{Mpc}^{-1}$. The value of the late time clustering amplitude ($\se$) is skewed towards the value, $\se = 0.77^{+0.04}_{-0.03} $, as obtained by the galaxy cluster counts using thermal Sunyaev-Zel'dovich (tSZ) signature~\cite{planck2018cp,zubeldia2019mnras}. Thus, \Planck data alone alleviates the $\se$ tension in the \modellate. Figure \ref{im:tri1} highlights the positive correlation between $H_{0}$ and $\se$ and strong negative correlations of $\Omega_{m}$ with $H_{0}$ and $\se$. The parameter $w_{0}$ has negative correlations with $w_{1}$, $H_{0}$ and $\se$ and positive correlation with $\Omega_{m}$. The coupling parameter ($\beta_{0}$) is uncorrelated to the others. 

Addition of the BAO to the \Planck data, increases the value of $\beta_{0}$ to $0.00814^{+0.00755+0.0146}_{-0.00755-0.0151}$ with zero outside the $1\sigma$ region. The \Planck and BAO combination cannot constrain the parameters $w_{0}$ and $w_{1}$. The mean value of the Hubble parameter increases considerably but is still smaller than the corresponding value for \lcdm, $H_{0} = 67.66\pm 0.42$ in $\mbox{km s}^{-1} \mbox{Mpc}^{-1}$ in the $1\sigma$ region. The considerable decrease in error bar increased the $H_{0}$ tension to $\sim 4\sigma$. The values of $\Omega_{m}$ decreases and $\se$ increases and are higher than the \lcdm counterpart ($\Omega_{m} = 0.3111\pm 0.0056$ and $\se = 0.8102\pm 0.006$) in the $1\sigma$ region. Thus, addition of the BAO data to the \Planck data restores the $\se$ tension ($\sim 0.79\sigma$) in \modellate. The combination also lowers the error regions substantially. 

Interestingly, addition of $\fsg$ to the \Planck data changes the parameter mean values in the similar fashion like the \Planck and BAO combination but the error bars become higher. This is also clear from Fig.\ \ref{im:tri1}. The mean value of $\beta_{0} (=0.00752^{+0.00757+0.0145}_{-0.00757-0.0151})$ is smaller the \Planck and BAO combination. Clearly, addition of the $\fsg$ data restores the $\se$ tension in \modellate.

Addition of the BAO and Pantheon to the \Planck data, increases the mean value of $\beta_{0} (=0.00859^{+0.00745+0.0145}_{-0.00745-0.0148} )$ with zero in the $2\sigma$ region. The parameters $w_{0}$ and $w_{1}$ still remain unconstrained. The combination increases the $H_{0}$ mean value but is still slightly smaller than the fiducial \lcdm value. The central value of $\Omega_{m}$ at the present epoch remains slightly larger whereas $\se$ remains slightly smaller than the \lcdm case. Clearly, the $\se$ tension is restored.

Combining $\fsg$ data with \Planck, BAO and Pantheon lowers the mean values of both $\Omega_{m}$ and $\se$ but increases the value of $H_{0}$ compared to the baseline \Planck values~\cite{planck2018cp}. Thus, addition of all the datasets worsen the $H_{0}$ tension ($\sim 4.2\sigma$) and the $\se$ tension ($\sim 0.87\sigma$). The mean value of $\beta_{0} (=0.00818^{+0.00731+0.0142}_{-0.00731-0.0146})$ decreases slightly with zero in the $2\sigma$ region. Although the constraints on $w_{0}$ and $w_{1}$ tightens, they still remain unconstrained.

Combination of all the datasets significantly reduced the error bars. The parameters, $\beta_{0}$ and $w_{1}$ become very weakly correlated with other parameters. However, the correlations among the rest of the parameters remain unchanged. 
\begin{center}
\begin{table*}[!htbp]
\centering
\caption{Observational constraints on the nine dependent model parameters with three derived parameters separated by a horizontal line and the error bars correspond to $68\%$ confidence level for {\bf Model L}, using different observational datasets.}
\label{tab:mean-1}
\begin{adjustbox}{width=0.9\textwidth}
\begin{tabular} { l  c c c c c}
\hline\hline \noalign{\vskip 2pt}
 {\normalsize Parameter} &  {\normalsize \Planck} &  {\normalsize \Planck \dataplus $f\sigma_{8}$} &  {\normalsize \Planck \dataplus BAO} &  {\normalsize \thead{\Planck \\ \dataplus BAO \dataplus Pantheon}} &  {\normalsize \thead{\Planck \dataplus BAO \\ \dataplus Pantheon \dataplus $f\sigma_{8}$}}\\
\hline \noalign{\vskip 2pt}
\rule[-1.5ex]{0pt}{2.7ex}{\boldmath$\Omega_b h^2   $} & $0.022362\pm 0.000168      $ & $0.022483\pm 0.000163      $ & $0.022487\pm 0.000156      $ & $0.022500\pm 0.000154      $ & $0.022542\pm 0.000152      $\\
\rule[-1.5ex]{0pt}{2.7ex}{\boldmath$\Omega_c h^2   $} & $0.12005\pm 0.00129        $ & $0.11853\pm 0.00117        $ & $0.11848\pm 0.00102        $ & $0.118381\pm 0.000977      $ & $0.117838\pm 0.000927      $\\
\rule[-1.5ex]{0pt}{2.7ex}{\boldmath$100\theta_{MC} $} & $1.040773\pm 0.000326      $ & $1.040938\pm 0.000316      $ & $1.040951\pm 0.000315      $ & $1.040954\pm 0.000316      $ & $1.041007\pm 0.000313      $\\
\rule[-1.5ex]{0pt}{2.7ex}{\boldmath$\tau           $} & $0.05475\pm 0.00773        $ & $0.05641^{+0.00705}_{-0.00794}$ & $0.05732^{+0.00701}_{-0.00787}$ & $0.05707^{+0.00691}_{-0.00777}$ & $0.05791\pm 0.00760        $\\
\rule[-1.5ex]{0pt}{2.7ex}{\boldmath$\beta_0        $} & $0.00788\pm 0.00815        $ & $0.00752\pm 0.00757        $ & $0.00814\pm 0.00755        $ & $0.00859\pm 0.00745        $ & $0.00818\pm 0.00731        $\\
\rule[-1.5ex]{0pt}{2.7ex}{\boldmath$w_0            $} & $< -0.909                  $ & $< -0.976                  $ & $< -0.968                  $ & $< -0.980                  $ & $< -0.985                  $\\
\rule[-1.5ex]{0pt}{2.7ex}{\boldmath$w_{1}          $} & $< 0.174                   $ & $< 0.0672                  $ & $< 0.0707                  $ & $< 0.0623                  $ & $< 0.0500                  $\\
\rule[-1.5ex]{0pt}{2.7ex}{\boldmath${\rm{ln}}(10^{10} A_s)$} & $3.0489\pm 0.0149          $ & $3.0489\pm 0.0147          $ & $3.0512\pm 0.0148          $ & $3.0507\pm 0.0145          $ & $3.0512\pm 0.0146          $\\
\rule[-1.5ex]{0pt}{2.7ex}{\boldmath$n_s            $} & $0.96330\pm 0.00444        $ & $0.96672\pm 0.00436        $ & $0.96674\pm 0.00413        $ & $0.96690\pm 0.00418        $ & $0.96818\pm 0.00412        $\\
\hline \noalign{\vskip 2pt}
\rule[-1.5ex]{0pt}{2.7ex}$H_0 \left[\mbox{km s}^{-1} \mbox{Mpc}^{-1}\right]                       $ & $63.98^{+2.45}_{-1.47}     $ & $66.93^{+1.04}_{-0.719}    $ & $66.770^{+0.792}_{-0.602}  $ & $67.179^{+0.579}_{-0.521}  $ & $67.596\pm 0.524           $\\
\rule[-1.5ex]{0pt}{2.7ex}$\Omega_m                  $ & $0.3507^{+0.0157}_{-0.0292}$ & $0.31643^{+0.00846}_{-0.0116}$ & $0.31778^{+0.00681}_{-0.00832}$ & $0.31368\pm 0.00632        $ & $0.30871\pm 0.00598        $\\
\rule[-1.5ex]{0pt}{2.7ex}$\sigma_8                  $ & $0.7825^{+0.0228}_{-0.0141}$ & $0.8027^{+0.0102}_{-0.00836}$ & $0.8020^{+0.0104}_{-0.00892}$ & $0.80541\pm 0.00862        $ & $0.80560\pm 0.00803        $\\
\hline\hline
\end{tabular}
\end{adjustbox}
\end{table*}
\begin{figure*}[!htbp]
        \centering
         \includegraphics[width=0.75\linewidth]{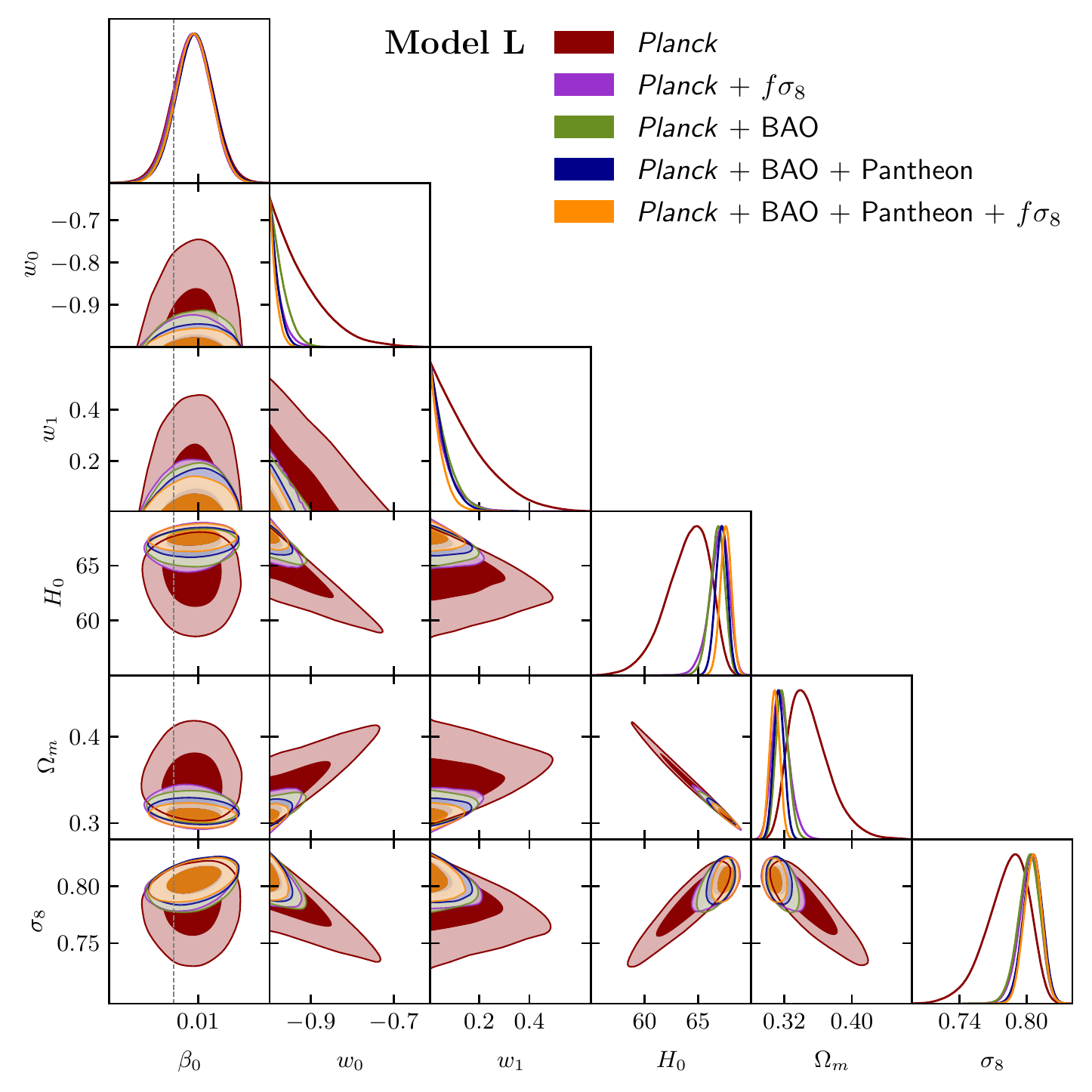}
         \caption{Plot of 1-dimensional marginalised posterior distributions and 2-dimensional marginalised constraint contours on the parameters of {\bf Model L} containing $68\%$ and $95\%$ probability. The dashed line represents the $\beta_{0}=0$ value.}\label{im:tri1}
\end{figure*}
\end{center}

\subsection{Model E}
For \modelearly, the mean values with $1\sigma$ errors of the nine free parameters along with the three derived parameters, $H_{0}$, $\Omega_{m}$ and $\se$, are given in Table \ref{tab:mean-2}. The correlations of the model parameters ($\beta_{0}$, $w_{0}$, $w_{1}$) with the derived parameters ($H_{0}$, $\Omega_{m}$, $\se$) and their marginalised contours are shown in Fig.\ \ref{im:tri2}. When only \Planck data is considered, the mean value of $\beta_{0}(=0.0339^{+0.0372+0.0724}_{-0.0372-0.0746})$ is large compared to that in \modellate, though $\beta_{0}=0$ remains within the $1\sigma$ region. The CPL parameters, $w_{0} (< -0.914< -0.809)$ and $w_{1} (< 0.168 < 0.355)$, remain unconstrained even within the $2\sigma$ region. The values of $H_{0}$ is greater and that of $\se$ is slightly grater whereas $\Omega_{m}$ is slightly lower than those in \modellate. Similar to \modellate, the discrepancy in the value of $H_{0}$ with the local measurement is at $3.5\sigma$. In \modelearly also, the \Planck data alleviates the $\se$ tension. The distinguishing feature of \modelearly is that the mean value of $\beta_{0}$ is greater than that obtained in \modellate for all the dataset combinations. 

Addition of the BAO to the \Planck data, increases the mean value of $\beta_{0}(=0.0432^{+0.0376+0.0733}_{-0.0376-0.0744})$ with zero allowed in the $2\sigma$ region. The parameters $w_{0}$ and $w_{1}$ remain unconstrained. The mean value of $H_{0}$ increases considerably but is still smaller than the corresponding value for \lcdm. The values of $\Omega_{m}$ decreases and $\se$ increases and are higher than the \lcdm counterpart. The addition of BAO data to \Planck data restores the $H_{0}$ ($\sim 4\sigma$) and $\se$ ($\sim 0.77\sigma$) tensions in \modelearly. The combination also lowers the error bars considerably.

Similar to \modellate, addition of $\fsg$ to the \Planck data changes the parameter mean values like the \Planck \dataplus BAO combination but the error bars still remain a little higher. This is also clear from Fig.\ \ref{im:tri2}. The mean value of $\beta_{0} (=0.0395^{+0.0381+0.0735}_{-0.0381-0.0750})$ is slightly smaller than the \Planck \dataplus BAO combination. The $H_{0}$ and $\se$ tensions are restored on addition of $\fsg$ to the \Planck data.

Combining BAO and Pantheon with \Planck data increases the mean value of $\beta_{0} (=0.0448^{+0.0377+0.0738}_{-0.0377-0.0733})$ and $\beta_{0} =0$ is within the $2\sigma$ region. The \Planck \dataplus BAO \dataplus Pantheon results in a very small change in the mean values of the parameters along with reduced error bars. The mean values of $H_{0}$ and $\se$ increase and $\Omega_{m}$ decreases relative to the \Planck \dataplus BAO combination. Again, the $\se$ tensions are not alleviated.

Addition of $\fsg$ to the combination \Planck \dataplus BAO \dataplus Pantheon, increases the mean value of $H_{0}$ slightly and decreases the mean value of $\Omega_{m}$ very slightly keeping $\se$ almost unchanged. The mean value of $\beta_{0} (=0.0446^{+0.0370+0.0724}_{-0.0370-0.0726})$ decreases slightly with zero in the $2\sigma$ region. Clearly, the addition of datasets do not improve the $H_{0}$ and $\se$ tension in \modelearly. Addition of the datasets significantly reduces the error bars. The correlations between the parameters for \modelearly remain same as in \modellate. 
\begin{center}
\begin{table*}[!htbp]
\centering
\caption{Observational constraints on the nine dependent model parameters with three derived parameters separated by a horizontal line and the error bars correspond to $68\%$ confidence level for {\bf Model E}, using different observational datasets.}
\label{tab:mean-2}
\begin{adjustbox}{width=0.9\textwidth}
\begin{tabular} { l  c c c c c}
\hline\hline \noalign{\vskip 2pt}
 {\normalsize Parameter} &  {\normalsize \Planck} &  {\normalsize \Planck \dataplus $f\sigma_{8}$} &  {\normalsize \Planck \dataplus BAO} &  {\normalsize \thead{\Planck \\ \dataplus BAO \dataplus Pantheon}} &  {\normalsize \thead{\Planck \dataplus BAO \\ \dataplus Pantheon \dataplus $f\sigma_{8}$}}\\
\hline \noalign{\vskip 2pt}
\rule[-1.5ex]{0pt}{2.7ex}{\boldmath$\Omega_b h^2   $} & $0.022358\pm 0.000165      $ & $0.022490\pm 0.000162      $ & $0.022489\pm 0.000156      $ & $0.022500\pm 0.000152      $ & $0.022546\pm 0.000151      $\\
\rule[-1.5ex]{0pt}{2.7ex}{\boldmath$\Omega_c h^2   $} & $0.12008\pm 0.00126        $ & $0.11848\pm 0.00117        $ & $0.11850\pm 0.00101        $ & $0.118405\pm 0.000970      $ & $0.117845\pm 0.000909      $\\
\rule[-1.5ex]{0pt}{2.7ex}{\boldmath$100\theta_{MC} $} & $1.040769\pm 0.000324      $ & $1.040941\pm 0.000318      $ & $1.040941\pm 0.000313      $ & $1.040945\pm 0.000315      $ & $1.040999\pm 0.000313      $\\
\rule[-1.5ex]{0pt}{2.7ex}{\boldmath$\tau           $} & $0.05466^{+0.00699}_{-0.00779}$ & $0.05630^{+0.00703}_{-0.00797}$ & $0.05704^{+0.00704}_{-0.00792}$ & $0.05697\pm 0.00749        $ & $0.05778^{+0.00700}_{-0.00790}$\\
\rule[-1.5ex]{0pt}{2.7ex}{\boldmath$\beta_0        $} & $0.0339\pm 0.0372          $ & $0.0395\pm 0.0381          $ & $0.0432\pm 0.0376          $ & $0.0448\pm 0.0377          $ & $0.0446\pm 0.0370          $\\
\rule[-1.5ex]{0pt}{2.7ex}{\boldmath$w_0            $} & $< -0.914                  $ & $< -0.977                  $ & $< -0.969                  $ & $< -0.981                  $ & $< -0.985                  $\\
\rule[-1.5ex]{0pt}{2.7ex}{\boldmath$w_{1}          $} & $< 0.168                   $ & $< 0.0645                  $ & $< 0.0707                  $ & $< 0.0604                  $ & $< 0.0489                  $\\
\rule[-1.5ex]{0pt}{2.7ex}{\boldmath${\rm{ln}}(10^{10} A_s)$} & $3.0486\pm 0.0147          $ & $3.0488\pm 0.0148          $ & $3.0509\pm 0.0148          $ & $3.0507\pm 0.0144          $ & $3.0511\pm 0.0146          $\\
\rule[-1.5ex]{0pt}{2.7ex}{\boldmath$n_s            $} & $0.96315\pm 0.00453        $ & $0.96681\pm 0.00434        $ & $0.96652\pm 0.00419        $ & $0.96672\pm 0.00418        $ & $0.96802\pm 0.00404        $\\
\hline \noalign{\vskip 2pt}
\rule[-1.5ex]{0pt}{2.7ex}$H_0 \left[\mbox{km s}^{-1} \mbox{Mpc}^{-1}\right]                       $ & $64.12^{+2.40}_{-1.39}     $ & $67.00^{+1.02}_{-0.702}    $ & $66.787^{+0.775}_{-0.600}  $ & $67.200^{+0.577}_{-0.516}  $ & $67.631\pm 0.516           $\\
\rule[-1.5ex]{0pt}{2.7ex}$\Omega_m                  $ & $0.3492^{+0.0149}_{-0.0282}$ & $0.31569^{+0.00834}_{-0.0114}$ & $0.31765^{+0.00678}_{-0.00812}$ & $0.31353^{+0.00590}_{-0.00658}$ & $0.30842\pm 0.00588        $\\
\rule[-1.5ex]{0pt}{2.7ex}$\sigma_8                  $ & $0.7836^{+0.0221}_{-0.0138}$ & $0.80265^{+0.00992}_{-0.00800}$ & $0.8019^{+0.0102}_{-0.00866}$ & $0.80539\pm 0.00830        $ & $0.80573\pm 0.00774        $\\
\hline\hline
\end{tabular}
\end{adjustbox}
\end{table*}
\begin{figure*}[!htbp]
        \centering
         \includegraphics[width=0.75\linewidth]{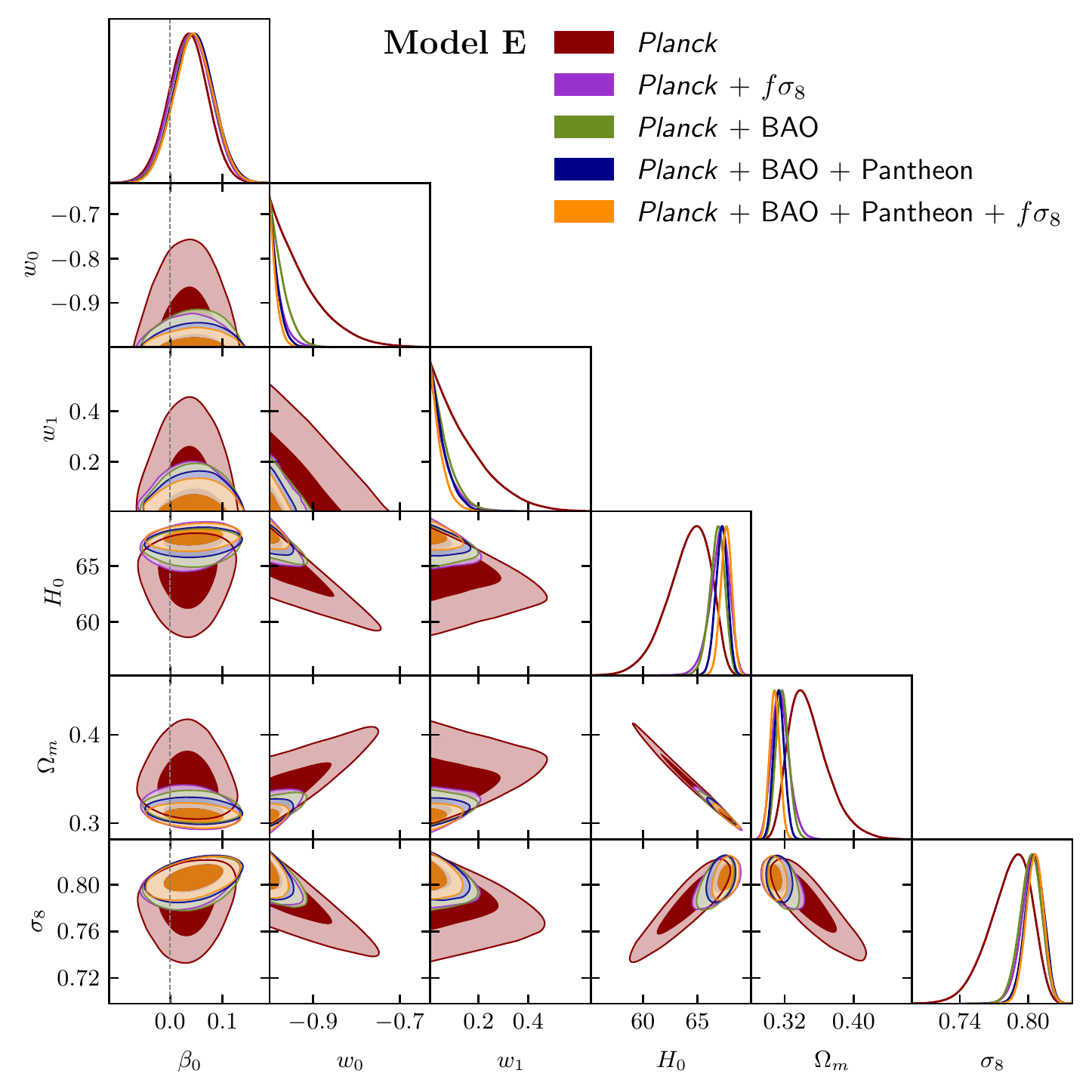}
         \caption{Plot of 1-dimensional marginalised posterior distributions and 2-dimensional marginalised constraint contours on the parameters of {\bf Model E} containing $68\%$ and $95\%$ probability. The dashed line represents the $\beta_{0}=0$ value.}\label{im:tri2}
\end{figure*}
\end{center}

\subsection{Model C}
The mean values of the parameters with $1\sigma$ errors for \modelcons are given in Table \ref{tab:mean-3}. In Table \ref{tab:mean-3}, the mean values and $1\sigma$ errors of the three derived parameters, $H_{0}$, $\Omega_{m}$ and $\se$, are also quoted. The correlations between the model parameters ($\beta_{0}$, $w_{0}$, $w_{1}$) and the derived parameters ($H_{0}$, $\Omega_{m}$, $\se$) along with their marginalised contours are shown in Fig.\ \ref{im:tri3}. The parameter values of \modelcons are very close to those of \modellate and they respond to the datasets in the similar fashion as well. Similar to \modellate and \modelearly, $H_{0}$ tension is at $\sim4\sigma$ and only the \Planck data alleviates the $\se$ tension in \modelcons whereas consideration of other datasets restore the tension. The main difference is that the mean values of $\beta_{0}$ in \modelcons is slightly smaller than that in \modellate. These features are clearly seen from Table \ref{tab:mean-3}.
\begin{center}
\begin{table*}[!htbp]
\centering
\caption{Observational constraints on the nine dependent model parameters with three derived parameters separated by a horizontal line and the error bars correspond to $68\%$ confidence level for {\bf Model C}, using different observational datasets.}
\label{tab:mean-3}
\begin{adjustbox}{width=0.9\textwidth}
\begin{tabular} { l  c c c c c}
\hline\hline \noalign{\vskip 2pt}
 {\normalsize Parameter} &  {\normalsize \Planck} &  {\normalsize \Planck \dataplus $f\sigma_{8}$} &  {\normalsize \Planck \dataplus BAO} &  {\normalsize \thead{\Planck \\ \dataplus BAO \dataplus Pantheon}} &  {\normalsize \thead{\Planck \dataplus BAO \\ \dataplus Pantheon \dataplus $f\sigma_{8}$}}\\
\hline \noalign{\vskip 2pt}
\rule[-1.5ex]{0pt}{2.7ex}{\boldmath$\Omega_b h^2   $} & $0.022358\pm 0.000164      $ & $0.022482\pm 0.000164      $ & $0.022487\pm 0.000156      $ & $0.022499\pm 0.000151      $ & $0.022545\pm 0.000152      $\\
\rule[-1.5ex]{0pt}{2.7ex}{\boldmath$\Omega_c h^2   $} & $0.12007\pm 0.00128        $ & $0.11854\pm 0.00118        $ & $0.11849\pm 0.00100        $ & $0.118388\pm 0.000977      $ & $0.117824\pm 0.000935      $\\
\rule[-1.5ex]{0pt}{2.7ex}{\boldmath$100\theta_{MC} $} & $1.040772\pm 0.000322      $ & $1.040939\pm 0.000321      $ & $1.040947\pm 0.000314      $ & $1.040954\pm 0.000312      $ & $1.041011\pm 0.000311      $\\
\rule[-1.5ex]{0pt}{2.7ex}{\boldmath$\tau           $} & $0.05491\pm 0.00757        $ & $0.05638^{+0.00708}_{-0.00788}$ & $0.05718^{+0.00685}_{-0.00788}$ & $0.05730\pm 0.00751        $ & $0.05800^{+0.00707}_{-0.00790}$\\
\rule[-1.5ex]{0pt}{2.7ex}{\boldmath$\beta_0        $} & $0.00624\pm 0.00673        $ & $0.00621\pm 0.00626        $ & $0.00696\pm 0.00629        $ & $0.00708\pm 0.00631        $ & $0.00693\pm 0.00615        $\\
\rule[-1.5ex]{0pt}{2.7ex}{\boldmath$w_0            $} & $< -0.907                  $ & $< -0.977                  $ & $< -0.969                  $ & $< -0.981                  $ & $< -0.985                  $\\
\rule[-1.5ex]{0pt}{2.7ex}{\boldmath$w_{1}          $} & $< 0.174                   $ & $< 0.0681                  $ & $< 0.0728                  $ & $< 0.0610                  $ & $< 0.0511                  $\\
\rule[-1.5ex]{0pt}{2.7ex}{\boldmath${\rm{ln}}(10^{10} A_s)$} & $3.0493\pm 0.0146          $ & $3.0489\pm 0.0147          $ & $3.0511\pm 0.0146          $ & $3.0511\pm 0.0144          $ & $3.0514\pm 0.0147          $\\
\rule[-1.5ex]{0pt}{2.7ex}{\boldmath$n_s            $} & $0.96331\pm 0.00444        $ & $0.96670\pm 0.00435        $ & $0.96670\pm 0.00416        $ & $0.96695\pm 0.00415        $ & $0.96823\pm 0.00414        $\\
\hline \noalign{\vskip 2pt}
\rule[-1.5ex]{0pt}{2.7ex}$H_0 \left[\mbox{km s}^{-1} \mbox{Mpc}^{-1}\right]                      $ & $63.93^{+2.51}_{-1.44}     $ & $66.93^{+1.03}_{-0.714}    $ & $66.759^{+0.795}_{-0.594}  $ & $67.187^{+0.581}_{-0.516}  $ & $67.606^{+0.552}_{-0.493}  $\\
\rule[-1.5ex]{0pt}{2.7ex}$\Omega_m                  $ & $0.3513^{+0.0153}_{-0.0299}$ & $0.31643^{+0.00843}_{-0.0115}$ & $0.31789^{+0.00671}_{-0.00832}$ & $0.31361\pm 0.00631        $ & $0.30860\pm 0.00605        $\\
\rule[-1.5ex]{0pt}{2.7ex}$\sigma_8                  $ & $0.7821^{+0.0232}_{-0.0140}$ & $0.8026^{+0.0100}_{-0.00823}$ & $0.8020^{+0.0104}_{-0.00879}$ & $0.80558\pm 0.00856        $ & $0.80564\pm 0.00807        $\\
\hline\hline
\end{tabular}
\end{adjustbox}
\end{table*}
\begin{figure*}[!htbp]
        \centering
         \includegraphics[width=0.75\linewidth]{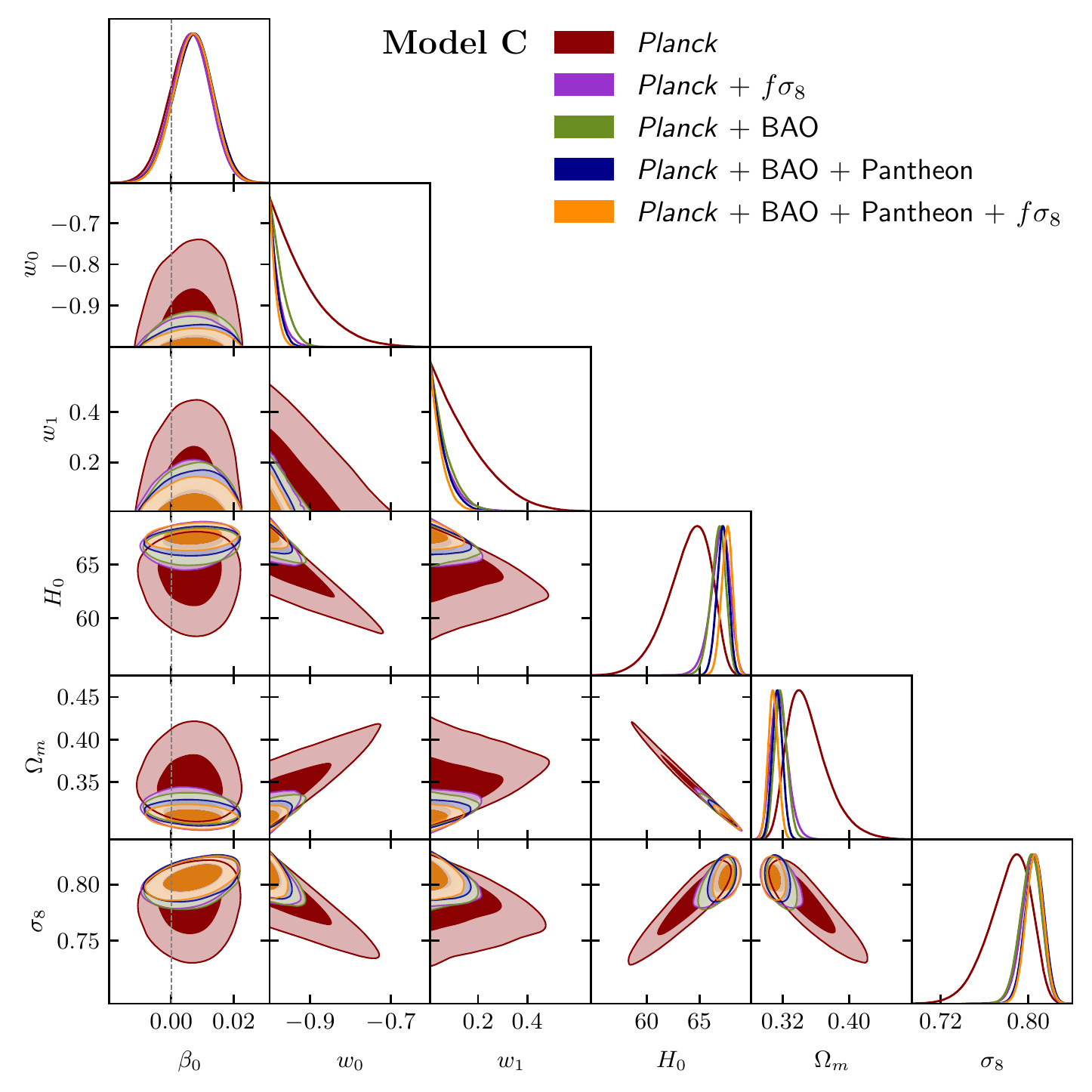}
         \caption{Plot of 1-dimensional marginalised posterior distributions and 2-dimensional marginalised constraint contours on the parameters of {\bf Model C} containing $68\%$ and $95\%$ probability. The dashed line represents the $\beta_{0}=0$ value.}\label{im:tri3}
\end{figure*}
\end{center}
%
\paragraph*{Tension in the $H_{0}$ parameter:}
Dark energy interacting with dark matter is a plausible scenario to resolve the $H_{0}$ tension. When $Q>0$, energy flows from dark matter to dark energy, and we have  less dark matter in the present epoch than in the uncoupled case. The locations and heights of the peaks and troughs in the CMB anisotropy depend on the combination $\Omega_{m} h^{2}$ of the present epoch. Since baryon energy density remains unaffected by the interaction, any change in the matter (which is a combination of baryonic matter and dark matter) density comes from the dark matter density. For low matter density with fixed $H_{0}$, the heights of the CMB peaks are higher (for details we refer to~\cite{dodelson2003}). So a model with less $\Omega_{c}$ at the present epoch will need to have a higher value of $H_{0}$ so as to keep the heights of the CMB peaks unaltered. Thus a larger value of $H_{0}$ will be obtained when constrained with the CMB data, which will reconcile the $H_{0}$ tension. It may be noted that in the present work, the uncoupled case is not the \lcdm model and $H_{0}$ tension is not alleviated.
%
%
\section{Bayesian Evidence}\label{sec:evi}
Finally, we aim to investigate which one of \modellate, \modelearly and \modelcons is statistically favoured by the observational data. The Bayesian evidence or more precisely, the logarithm of the Bayes factor, $\ln B_{ij}$ given in Eqn.\ (\ref{eq:bayesfac}), for each of the three models is calculated. Here, $i$ corresponds to \modellate, \modelearly and \modelcons for each of the the dataset combination, $j$ corresponds to the reference model, $M_{j}$. The details on the Bayes factor, $B_{ij}$, are discussed in Appendix \ref{sec:model-selection}. The fiducial \lcdm model is considered to be the reference model, and therefore, a negative value ($\ln B_{ij} < 0$) indicates a preference for the \lcdm model. The logarithmic Bayes factor, $\ln B_{ij}$, is calculated directly from the MCMC chains using the publicly available cosmological package \mcevidence\footnote{Available on GitHub: \href{https://github.com/yabebalFantaye/MCEvidence}{https://github.com/yabebalFantaye/MCEvidence}}~\cite{heavens2017prl, heavens2017ax}. The computed values of $\ln B_{ij}$ for \modellate, \modelearly and \modelcons are summarised in Table \ref{tab:evidence}. From Table \ref{tab:evidence}, it is clear that the \lcdm model is preferred over the interacting models by all the dataset combinations. However, the motive is to assess if there is any observationally preferable evolution stage when the interaction is significant. As can be seen from the relative differences of $\abs{\ln B_{ij}}$ (values corresponding to column $\Delta \ln B_{ij}$) in Table \ref{tab:evidence}), when compared with \modelcons, \modellate is strongly disfavoured while \modelearly is weakly disfavoured by observational data over \modelcons.
\begin{table}[!htbp]
\begin{center}
\caption{The values of $\ln B_{ij}$, where $j$ is the \lcdm model and $i$ is the interacting model. A negative sign indicates $M_{j}$ is favoured over $M_{i}$. The $\abs{\ln B_{ij}}$ values are compared with Table \ref{tab:bij-values}. The column $\Delta \ln B_{ij}$ corresponds to the comparison of {\bf Model L} and {\bf Model E} with {\bf Model C}.
  }\label{tab:evidence}
\begin{tabular}{ccccc}
\hline \hline
\rule[-1ex]{0pt}{2.5ex}Model& Dataset&\hspace{0ex} ~~$\ln B_{ij}$&\hspace{0.5ex}~~~$\Delta \ln B_{ij}$\\
    \hline
    \multirow{5}{*}{\rule[-1ex]{0pt}{2.5ex}{\bf Model L}}& \Planck &\hspace{0ex} $-8.843$&\hspace{0.5ex} $-2.244$ \\
    \rule[-1ex]{0pt}{2.5ex}& \Planck \dataplus $f\sigma_{8}$ &\hspace{0ex} $-11.410$&\hspace{0.5ex} $-2.245$\\
    \rule[-1ex]{0pt}{2.5ex}& \Planck \dataplus BAO &\hspace{0ex} $-10.610$&\hspace{0.5ex} $-2.187$ \\
    \rule[-1ex]{0pt}{2.5ex}& \Planck \dataplus BAO \dataplus Pantheon &\hspace{0ex} $-11.354$&\hspace{0.5ex} $-2.104$\\
    \rule[-1ex]{0pt}{2.5ex}& \Planck \dataplus BAO \dataplus Pantheon \dataplus $f\sigma_{8}$ &\hspace{0ex}$-11.977$&\hspace{0.5ex} $-2.328$ \\
    \hline
    \multirow{5}{*}{\rule[-1ex]{0pt}{2.5ex}{\bf Model E}}& \Planck &\hspace{0ex} $-7.233$  &\hspace{0.5ex} $-0.633$\\
    \rule[-1ex]{0pt}{2.5ex}& \Planck \dataplus $f\sigma_{8}$ &\hspace{0ex} $-9.730$&\hspace{0.5ex} $-0.566$\\
    \rule[-1ex]{0pt}{2.5ex}& \Planck \dataplus BAO &\hspace{0ex} $-9.047$&\hspace{0.5ex} $-0.624$ \\
    \rule[-1ex]{0pt}{2.5ex}& \Planck \dataplus BAO \dataplus Pantheon &\hspace{0ex} $ -9.733$&\hspace{0.5ex} $-0.483$ \\
    \rule[-1ex]{0pt}{2.5ex}& \Planck \dataplus BAO \dataplus Pantheon \dataplus $f\sigma_{8}$ &\hspace{0ex} $-10.192$&\hspace{0.5ex} $-0.542$\\
    \hline
    \multirow{5}{*}{\rule[-1ex]{0pt}{2.5ex}{\bf Model C}}& \Planck &\hspace{0ex} $-6.599$ &\hspace{0.5ex} 0.0\\
    \rule[-1ex]{0pt}{2.5ex}& \Planck \dataplus $f\sigma_{8}$ &\hspace{0ex} $-9.164$&\hspace{0.5ex} 0.0 \\
    \rule[-1ex]{0pt}{2.5ex}&  \Planck \dataplus BAO &\hspace{0ex} $-8.423$&\hspace{0.5ex} 0.0\\
    \rule[-1ex]{0pt}{2.5ex}& \Planck \dataplus BAO \dataplus Pantheon &\hspace{0ex} $-9.250$&\hspace{0.5ex} 0.0\\
    \rule[-1ex]{0pt}{2.5ex}& \Planck \dataplus BAO \dataplus Pantheon \dataplus $f\sigma_{8}$ &\hspace{0ex} $-9.650$&\hspace{0.5ex} 0.0\\
\hline
\hline
\end{tabular}
\end{center}
\end{table}
%
%

\section{Summary and Discussion} \label{sec:sum}
The present work deals with the matter perturbations in a cosmological model where the dark energy has an interaction with the dark matter. We investigate the possibility whether the coupling parameter between the two dark components can evolve. We have considered two new examples, (a) the interaction is a recent phenomenon (\modellate; Eqn.\ (\ref{eq:b1})), and (b) the interaction is an early phenomenon (\modelearly; Eqn.\ (\ref{eq:b2})) and compared them with the normally talked about model where the coupling is a constant (\modelcons; Eqn.\ (\ref{eq:b3})), in the context of density perturbations. The results are compared with the standard \lcdm model as well. The rate of energy transfer is proportional to the dark energy density, $\rde$ and energy flows from dark matter to dark energy. The interaction term is given by Eqn.\ (\ref{eq:q1}). We have also considered the dark energy to have a dynamical EoS parameter, $\wde$ being given by the CPL parametrisation (Eqn.\ (\ref{eq:w-cpl})). 

We have worked out a detailed perturbation analysis of the models in the synchronous gauge and compared them with each other. The background dynamics of the three interacting models are almost the same, which is evident from the smallness of the coupling parameter and the domination of dark energy at late times. The signature of the presence of interaction at different epochs for different couplings are noticeable in the perturbation analysis.

In all the three interacting models, the fractional density perturbation of dark matter is marginally higher than that in a \lcdm model, indicating more clumping of matter. From the CMB temperature spectrum, matter power spectrum and the evolution of growth rate, we note that the presence of interaction for a brief period in the evolutionary history (\modelearly), makes the Universe behave like the \lcdm model with a slightly higher value of $\fsg$ at the epoch when the interaction prevails. The first part of the present work shows that \modelearly behaves in a closely similar fashion as the \lcdm model and leads to the conclusion that \modelearly performs better than \modellate and \modelcons in describing the evolutionary history of the Universe. 

To determine further the evolution stage when the interaction is significant, we have tested the interacting models with the observational datasets. We have tested \modelearly, \modellate and \modelcons against the recent observational datasets like CMB, BAO, Pantheon and RSD with the standard six parameters of \lcdm model along with the three model parameters, $\beta_{0}$, $w_{0}$ and $w_{1}$. We have obtained the mean value of the coupling parameter, $\beta_{0}$ to be positive, indicating an energy flow from dark matter to dark energy. When only CMB data is used, $\beta_{0}=0$ lies within the $1\sigma$ error region while when different combinations of the datasets are used, $\beta_{0}=0$ lies outside the $1\sigma$ error region. The priors of $w_{0}$ and $w_{1}$ are set such that $\wde$ remains in the quintessence region. Hence, $w_{0}$ and $w_{1}$ remain unconstrained. Moreover, for all the three interacting models, the $\se$ tension is alleviated when CMB data is used.  Though the estimated parameter values are prior dependent, it can be said conclusively that the CMB data and RSD data are in agreement when the interacting models are considered. Addition of other datasets restore the $\se$ tension in all the three interacting models. However, the tension in $H_{0}$ value persists for all the three interacting models.

From the Bayesian evidence analysis, we see that all the three interacting dark energy models are rejected by observational data when compared with the fiducial \lcdm model. However, a close scrutiny reveals that both \modelearly and \modelcons are favoured over \modellate. Though the Bayesian evidence analysis ever so slightly favours \modelcons over \modelearly, the difference is too small to choose a clear winner. Thus, to conclude from the results of the perturbation analysis and observational data we infer that the interaction, if present, is likely to be significant only at some early stage of evolution of the Universe.

\section*{Acknowledgement}

The author is indebted to Narayan Banerjee for valuable suggestions and discussions. The author would also like to thank Tuhin Ghosh, Supriya Pan and Ankan Mukherjee for their insightful comments and suggestions.


\appendix
\numberwithin{equation}{section}
\numberwithin{table}{section}
\section{Observational data and methodology}\label{sec:app}
Different observational datasets obtained from the publicly available cosmological probes have been used to constrain the parameters of the interacting models. The datasets used in this work are listed below.
\begin{description}
\item[CMB] 
We considered the cosmic microwave background (CMB) anisotropies data from the latest 2018 data release of the \Planck collaboration\footnote{Available at: \href{http://pla.esac.esa.int/pla/\#home}{ https://pla.esac.esa.int}}~\cite{planck2019cmb,planck2018cp}. The CMB likelihood consists of the low-$\ell$ temperature likelihood, $C_{\ell}^{TT}$, the low-$\ell$ polarization likelihood, $C_{\ell}^{EE}$, high-$\ell$ temperature-polarization likelihood, $C_{\ell}^{TE}$, high-$\ell$ combined TT, TE and EE  likelihood. The low-$\ell$ likelihoods span from $2\le \ell \le 29$ and the high-$\ell$ likelihoods consists of multipole values $\ell \ge 30$ and collectively make the combination \planckall. For CMB lensing data, the power spectrum of the lensing potential measured by \Planck collaboration is used. The \planckall, along with the lensing likelihood (\plancklensing) are denoted as `\Planck' in the results given in Sect.\ \ref{sec:mcmc}. References~\cite{planck2019cmb,planck2018cp} provide a detailed study of the CMB likelihoods.
\item[BAO]
The photon-baryon fluid fluctuations in the early Universe leave their signatures as the acoustic peaks in the CMB anisotropies power spectrum. The anisotropies of baryon acoustic oscillations (BAO) provide tighter constraints on the cosmological parameters~\cite{eisenstein2005apj}. The BAO surveys measure the ratio, $D_{V}/r_{d}$ at different effective redshifts. The quantity $D_{V}$ is related to the comoving angular diameter $D_{M}$ and Hubble parameter $H$ as
\begin{equation}
D_{V}\paren*{z} = \left[D_{M}^{2}\paren*{z} \frac{c\,z}{H\paren*{z}}\right] ^{1/3},
\end{equation}
and $r_{d}$ refers to the comoving sound horizon at the end of baryon drag epoch. For the BAO data, three surveys are considered: the 6dF Galaxy Survey (6dFGS) measurements~\cite{beutler2011mnras} at redshift $z = 0.106$, the Main Galaxy Sample of Data Release $7$ of the Sloan Digital Sky Survey (SDSS-MGS)~\cite{ross2015mnras} at redshift $z = 0.15$ and the latest Data Release $12$ (DR12) of the Baryon Oscillation Spectroscopic Survey (BOSS) of the Sloan Digital Sky Survey (SDSS) III at redshifts $z = 0.38$, $0.51$ and $0.61$~\cite{alam2017sdss3}.
\item[Pantheon] 
We considered the latest `Pantheon' catalogue for the luminosity distance measurements of the Type Ia supernovae (SNe Ia)~\cite{scolnic2018apj}. The Pantheon sample is the compilation of 276 supernovae discovered by the Pan-STARRS1 Medium Deep Survey at $0.03 < z < 0.65$ and various low redshift and Hubble Space Telescope (HST) samples to give a total of 1048 supernovae data in the redshift range $0.01 < z < 2.3$.
\item[RSD]
Redshift-space distortion (RSD) is the cosmological effect where spatial galaxy maps produced by measuring distances from the spectroscopic redshift surveys show an anisotropic galaxy distribution. These galaxy anisotropies arise due to the galaxy recession velocities having components from both the Hubble flow and comoving peculiar velocities from the motions of galaxies and result in the anisotropies of the observed power spectrum. However, additional anisotropies in the power spectrum arise due to incorrect fiducial cosmology, $H\paren*{z}$ while converting the relative redshifts to comoving coordinates. The introduction of anisotropies due to incorrect fiducial cosmology is called Alcock-Paczy{\'n}ski (AP) effect~\cite{alcock1979nature}. The RSD surveys measure the matter peculiar velocities and provide the galaxy matter density perturbation, $\delta_{g}$~\cite{kaiser1987mnras}. As mentioned in Sect.\ \ref{sec:result}, the combination $\fsg$ is the widely used quantity to study the growth rate of the matter density perturbation. In the present work, we considered the $\fsg$ data compilation by Nesseris \etal~\cite{nesseris2017prd}, Sagredo \etal~\cite{sagredo2018prd} and Skara and Perivolaropoulos~\cite{skara2020prd}. The surveys and the corresponding data points used in this work are shown in Table \ref{tab:fs8data}, along with the corresponding fiducial cosmology used by the collaborations to convert redshift to distance in each case. The fiducial cosmology in Table \ref{tab:fs8data} is used to correct the AP effect following Macaulay \etal~\cite{macaulay2013prl} as discussed in~\cite{sagredo2018prd,skara2020prd}. The RSD measurement is denoted as `$\fsg$' data in the results given in Sect.\ \ref{sec:mcmc}.
\end{description}
\begin{table*}[!htbp]
\centering
\caption{
A compilation of $\fsg$ measurements with redshift $z$ and fiducial value of $\Omega_{m}$ from different surveys.
}\label{tab:fs8data}
\begin{tabular}{cccccc}
\hline
\hline
\rule[-1ex]{0pt}{2.ex}Survey &\hspace{7ex} $z$ &\hspace{7ex} $f\sigma_8(z)$ &\hspace{7ex} $\Omega_m$&\hspace{7ex} Refs. \\
\hline
\rule[-1ex]{0pt}{2.ex}6dFGS+SnIa &\hspace{7ex} $0.02$ &\hspace{7ex}  $0.428\pm 0.0465$ &  \hspace{7ex}$0.3$& \hspace{7ex}~\cite{huterer2017jcap} \\ 
\rule[-1ex]{0pt}{2.ex}SnIa+IRAS &\hspace{7ex} $0.02$&\hspace{7ex} $0.398 \pm 0.065$ &\hspace{7ex}$0.3$&\hspace{7ex}~\cite{turnbull2012mnras},~\cite{hudson2012ajl}\\
\rule[-1ex]{0pt}{2.ex}2MASS &\hspace{7ex} $0.02$&\hspace{7ex} $0.314 \pm 0.048$ &\hspace{7ex} $0.266$&\hspace{7ex}~\cite{davis2011mnras},~\cite{hudson2012ajl} \\
\rule[-1ex]{0pt}{2.ex}SDSS-veloc &\hspace{7ex} $0.10$ &\hspace{7ex} $0.370\pm 0.130$ &\hspace{7ex} $0.3$&\hspace{7ex}~\cite{feix2015prl}  & \\
\rule[-1ex]{0pt}{2.ex}SDSS-MGS &\hspace{7ex} $0.15$ &\hspace{7ex} $0.490\pm0.145$ &\hspace{7ex} $0.31$&\hspace{7ex}~\cite{howlett2015mnras} \\
\rule[-1ex]{0pt}{2.ex}2dFGRS &\hspace{7ex} $0.17$ &\hspace{7ex} $0.510\pm 0.060$ &\hspace{7ex} $0.3$&\hspace{7ex}~\cite{song2009jcap} \\
\rule[-1ex]{0pt}{2.ex}GAMA &\hspace{7ex} $0.18$ &\hspace{7ex} $0.360\pm 0.090$ &\hspace{7ex} $0.27$&\hspace{7ex}~\cite{blake2013mnras}\\
\rule[-1ex]{0pt}{2.ex}GAMA &\hspace{7ex} $0.38$ &\hspace{7ex} $0.440\pm 0.060$ &&\hspace{7ex}~\cite{blake2013mnras} \\
\rule[-1ex]{0pt}{2.ex}SDSS-LRG-200 &\hspace{7ex} $0.25$ &\hspace{7ex} $0.3512\pm 0.0583$ &\hspace{7ex} $0.25$&\hspace{7ex}~\cite{samushia2012mnras}\\
\rule[-1ex]{0pt}{2.ex}SDSS-LRG-200 &\hspace{7ex} $0.37$ &\hspace{7ex} $0.4602\pm 0.0378$ &&\hspace{7ex}~\cite{samushia2012mnras}\\
\rule[-1ex]{0pt}{2.ex}BOSS-LOWZ&\hspace{7ex} $0.32$ &\hspace{7ex} $0.384\pm 0.095$ &\hspace{7ex} $0.274$&\hspace{7ex}~\cite{sanchez2014mnras} \\
\rule[-1ex]{0pt}{2.ex}SDSS-CMASS &\hspace{7ex} $0.59$ &\hspace{7ex} $0.488\pm 0.060$ &\hspace{7ex} $0.307115$&\hspace{7ex}~\cite{chuang2016mnras}\\
\rule[-1ex]{0pt}{2.ex}WiggleZ &\hspace{7ex} $0.44$ &\hspace{7ex} $0.413\pm 0.080$ &\hspace{7ex} $0.27$&\hspace{7ex}~\cite{blake2012mnras} \\
\rule[-1ex]{0pt}{2.ex}WiggleZ &\hspace{7ex} $0.60$ &\hspace{7ex} $0.390\pm 0.063$ &\hspace{7ex} $\mathbf{C}_{ij}\rightarrow$ Eq.~(\ref{eq:wigglez})&\hspace{7ex}~\cite{blake2012mnras} \\
\rule[-1ex]{0pt}{2.ex}WiggleZ &\hspace{7ex} $0.73$ &\hspace{7ex} $0.437\pm 0.072$ & &\hspace{7ex}~\cite{blake2012mnras}\\
\rule[-1ex]{0pt}{2.ex}VIPERS PDR-2&\hspace{7ex} $0.60$ &\hspace{7ex} $0.550\pm 0.120$ &\hspace{7ex} $0.3$&\hspace{7ex}~\cite{pezzotta2017aa} \\
\rule[-1ex]{0pt}{2.ex}VIPERS PDR-2&\hspace{7ex} $0.86$ &\hspace{7ex} $0.400\pm 0.110$ &&\hspace{7ex}~\cite{pezzotta2017aa}\\
\rule[-1ex]{0pt}{2.ex}FastSound&\hspace{7ex} $1.40$ &\hspace{7ex} $0.482\pm 0.116$ &\hspace{7ex} $0.27$&\hspace{7ex}~\cite{okumura2016pasj}\\
\rule[-1ex]{0pt}{2.ex}SDSS-IV&\hspace{7ex} $0.978$  &\hspace{7ex} $0.379 \pm 0.176$ &\hspace{7ex} $0.31$&\hspace{7ex}~\cite{zhao2018mnras}\\
\rule[-1ex]{0pt}{2.ex}SDSS-IV&\hspace{7ex} $1.23$ &\hspace{7ex} $ 0.385 \pm 0.099$ &\hspace{7ex} $\mathbf{C}_{ij}\rightarrow$ Eqn.~(\ref{eq:sdss})&\hspace{7ex}~\cite{zhao2018mnras}\\
\rule[-1ex]{0pt}{2.ex}SDSS-IV&\hspace{7ex} $1.526$ &\hspace{7ex} $0.342 \pm 0.070$  &&\hspace{7ex}~\cite{zhao2018mnras}\\
\rule[-1ex]{0pt}{2.ex}SDSS-IV&\hspace{7ex} $1.944$ &\hspace{7ex} $0.364 \pm 0.106$  &&\hspace{7ex}~\cite{zhao2018mnras}\\
\rule[-1ex]{0pt}{2.ex}VIPERS PDR2 &\hspace{7ex} $0.60$ &\hspace{7ex} $0.49 \pm 0.12$ &\hspace{7ex} $0.31$&\hspace{7ex}~\cite{mohammad2018aa}\\
\rule[-1ex]{0pt}{2.ex}VIPERS PDR2 &\hspace{7ex} $0.86$ &\hspace{7ex} $0.46 \pm 0.09$ &&\hspace{7ex}~\cite{mohammad2018aa}\\
\rule[-1ex]{0pt}{2.ex}BOSS DR12 voids &\hspace{7ex} $0.57$ &\hspace{7ex} $0.501 \pm 0.051$ &\hspace{7ex} $0.307$&\hspace{7ex}~\cite{nadathur2019prd}\\
\rule[-1ex]{0pt}{2.ex}2MTF 6dFGSv &\hspace{7ex} $0.03$ &\hspace{7ex} $0.404 \pm 0.0815$ &\hspace{7ex} $0.3121$&\hspace{7ex}~\cite{qin2019mnras}\\
\rule[-1ex]{0pt}{2.ex}SDSS-IV &\hspace{7ex} $0.72$ &\hspace{7ex} $0.454 \pm 0.139$ &\hspace{7ex} $0.31$&\hspace{7ex}~\cite{icaza2019mnras}\\
\hline \hline
\end{tabular}

\end{table*}
The covariance matrices of the data from the WiggleZ~\cite{blake2012mnras} and the SDSS-IV~\cite{zhao2018mnras} surveys are given as
\begin{equation} \label{eq:wigglez}
\mbox{\textbf{\large{C}}}_{\mbox{\scriptsize{WiggleZ}}}=10^{-3}
\left(
\begin{array}{ccc}
 6.400 & 2.570 & 0.000 \\
 2.570 & 3.969 & 2.540 \\
 0.000 & 2.540 & 5.184 \\
\end{array}
\right),
\end{equation}
and
\begin{equation} \label{eq:sdss}
\mbox{\textbf{\large{C}}}_{\mbox{\scriptsize{SDSS-IV}}}=10^{-2}
\left(
\begin{array}{cccc}
 3.098 & 0.892 & 0.329 & -0.021 \\
 0.892 & 0.980 & 0.436 & 0.076 \\
 0.329 & 0.436 & 0.490 & 0.350 \\
 -0.021 & 0.076 & 0.350 & 1.124 \\
\end{array}
\right)
\end{equation}
respectively.

To compare the interacting model with the observational data, we calculated the likelihood as
\begin{equation}\label{eq:chi-total}
\mathcal{L} \propto e^{-\chi^{2}/2}, \hspace{0.3cm} \mbox{where} \hspace{0.3cm} \chi^{2} = \chi_{\mbox{\scriptsize{CMB}}}^{2} +\chi_{\mbox{\scriptsize{BAO}}}^{2} + \chi_{\mbox{\scriptsize{Pantheon}}}^{2} +\chi_{\fsg}^{2}.
\end{equation}
According to Bayes theorem (see Eqn.\ \ref{eq:bayes-th}), the likelihood is the probability of the data given the model parameters. The quantity $\chi^{2}$ for any dataset is calculated as 
\begin{equation}
\chi^{2}_{i} = V^i\, \mathbf{C}_{ij}^{-1}\,V^j,
\end{equation}
where the vector, $V^{i}$ is written as
\begin{equation}
V^{i} = \Theta_{i}^{\scriptsize{\mbox{obs}}} - \Theta^{\scriptsize{\mbox{th}}}\paren*{z_{i},P}
\end{equation}
with $\Theta$ being the physical quantity corresponding to the observational data (\Planck, BAO, Pantheon, $\fsg$) used, $z_{i}$ being the corresponding redshift, $\mathbf{C}^{-1}_{ij}$ is the corresponding inverse covariance matrix and $P$ is the parameter space. The posterior distribution (see Eqn.\ \ref{eq:bayes-th}) is sampled using the Markov Chain Monte Carlo (MCMC) simulator through a suitably modified version of the publicly available code \cosmomc~\cite{lewis2013hha,lewis2002ah}. The statistical convergence of the MCMC chains for each model is set to satisfy the Gelman and Rubin criterion~\cite{gelman1992}, $R-1 \lesssim 0.01$. 

The correction for the Alcock-Paczy{\'n}ski effect is taken into account by the fiducial correction factor, $\mathcal{R}$~\cite{sagredo2018prd,skara2020prd} given as
\begin{equation} \label{eq:ratio}
\mathcal{R}(z)=\frac{H(z) D_A\paren*{z}}{H^{\scriptsize{\mbox{fid}}}\paren*{z} D^{\scriptsize{\mbox{fid}}}_A\paren*{z}}
\end{equation}
where $H\paren*{z}$ is the Hubble parameter and $D_A\paren*{z}$ is the angular diameter distance  of the interacting models and that of the fiducial cosmology are denoted with superscript `fid'. The corrected vector, $V_{\fsg}^i(z,P)$ is corrected as
\begin{equation} \label{eq:vector}
V_{\fsg}^i(z_{i},P) \equiv f\sigma_{8,i}^{\scriptsize{\mbox{obs}}} - \frac{f\sigma_{8}^{\scriptsize{\mbox{th}}}\paren*{z_{i},P}}{\mathcal{R}(z_{i})},
\end{equation}
where $f\sigma_{8,i}^{\scriptsize{\mbox{obs}}}$ is the $i$-th observed data point from Table \ref{tab:fs8data}, $f\sigma_{8}^{\scriptsize{\mbox{th}}}\paren*{z_{i},P}$ is the theoretical prediction at the same redshift $z_{i}$ and $P$ is the parameter vector given by Eqn.\ \ref{eq:parameter}. The corrected $\chi_{\fsg}^{2}$ is then written as
\begin{equation} \label{eq:chi-fs8}
\chi_{\fsg}^{2}= V_{\fsg}^i\, \mathbf{C}_{ij,\fsg}^{-1}\,V_{\fsg}^j,
\end{equation}
where $\mathbf{C}_{ij,\fsg}^{-1}$ is the inverse of the covariance matrix, $\mathbf{C}_{ij,\fsg}$ of the $\fsg$ dataset given by 
\begin{equation} \label{eq:totalcov}
\mathbf{C}_{ij,\fsg}=
\left(
\begin{array}{cccccc}
 \sigma_1^2 & 0 & \cdots & 0 & \cdots & 0 \\
 0 & \sigma_2^2 & \cdots & 0 & \cdots & 0 \\
 \vdots &  \vdots &  \vdots &  \vdots &  \vdots &  0 \\
 0 & 0 & \cdots & \mathbf{C}_{\mbox{\scriptsize{WiggleZ}}} & \cdots & 0 \\
 0 & 0 & \cdots & 0 & \mathbf{C}_{\mbox{\scriptsize{SDSS-IV}}} & 0 \\
0 & 0 & \cdots & 0 & \cdots & \sigma_{\scriptsize{N}}^2\\
\end{array}
\right)
\end{equation}
where $N=27$ corresponds to total number of data points in Table \ref{tab:fs8data}. Thus the covariance matrix, $\mathbf{C}_{ij,\fsg}$ is a $27 \times 27$ matrix with Eqns.\ (\ref{eq:wigglez}) and (\ref{eq:sdss}) at the positions of $\mathbf{C}_{\mbox{\scriptsize{WiggleZ}}}$ and $\mathbf{C}_{\mbox{\scriptsize{SDSS-IV}}}$ respectively and $\sigma_{i}$ is the error from Table \ref{tab:fs8data}. To use the RSD measurements, we added a new likelihood module to the publicly available \cosmomc package to calculate the corrected $\chi_{\fsg}^{2}$. The results obtained by analysing the MCMC chains are explained in Sect.\ \ref{sec:mcmc}.
%

\section{Model selection}\label{sec:model-selection}

Bayesian evidence is the Bayesian tool to compare models and is the integration of the likelihood over the multidimensional parameter space. Hence, it is also referred to as marginal likelihood. Using Bayes theorem, the posterior probability distribution of a model, $M$ with parameters $\Theta$ for the given particular dataset $x$ is obtained as
\begin{equation}\label{eq:bayes-th}
p\paren*{\Theta|x,M} = \frac{p\paren*{x|\Theta,M}\pi\paren*{\Theta|M}}{p\paren*{x|M}},
\end{equation}
where $p\paren*{x|\Theta,M}$ refers to the likelihood function, $\pi\paren*{\Theta|M}$ refers to the prior distribution and $p\paren*{x|M}$ refers to the Bayesian evidence. From Eqn.\ (\ref{eq:bayes-th}), the evidence follows as the integral over the unnormalised posterior distribution,
\begin{equation}
E \equiv p\paren*{x|M} = \int d\Theta p\paren*{x|\Theta,M}\pi\paren*{\Theta|M}.
\end{equation}
To compare model $M_{i}$ with the reference model $M_{j}$, the ratio of the evidences, called the Bayes factor is calculated.
\begin{equation}\label{eq:bayesfac}
B_{ij} = \frac{p\paren*{x|M_{i}}}{p\paren*{x|M_{j}}}.
\end{equation}
The calculation of the multidimensional integral is undoubtedly computationally expensive. This problem is solved by the method developed by Heavens \etal~\cite{heavens2017prl, heavens2017ax}, where the Bayesian evidence is estimated directly from the MCMC chains generated by \cosmomc. This method for evidence estimation is publicly available in the form of \mcevidence. The \mcevidence package provides with the logarithm of the Bayes factor, $\ln B_{ij}$. The value of $\ln B_{ij}$ is then used to assess if model $M_{i}$ is preferred over model $M_{j}$ and if so, what is the strength of preference, by using the revised Jeffreys scale (Table \ref{tab:bij-values}) by Kass and Raftery~\cite{kass1995jasa}. Thus, if $\ln B_{ij} > 0 $, model $M_{i}$ is preferred over model $M_{j}$.
\begin{table}[!htbp]
\begin{center}
\caption{Revised Jeffreys scale by Kass and Raftery to interpret the values of $\ln B_{ij}$ while comparing two models $M_{i}$ and $M_{j}$}\label{tab:bij-values}
\begin{tabular}{cc}
\hline \hline
\rule[-1ex]{0pt}{2.5ex}$\ln B_{ij}$ &  \hspace{24ex} Strength\\
\hline
\rule[-1ex]{0pt}{2.5ex}$0 \le \ln B_{ij} <1$&  \hspace{24ex} Weak \\
\rule[-1ex]{0pt}{2.5ex}$1 \le \ln B_{ij} <3$&  \hspace{24ex} Definite/Positive\\
\rule[-1ex]{0pt}{2.5ex}$3\le \ln B_{ij} <5$&  \hspace{24ex} Strong \\
\rule[-1ex]{0pt}{2.5ex}$\ln B_{ij} \ge 5$&  \hspace{24ex} Very strong \\
\hline \hline
\end{tabular}
\end{center}
\end{table}%

The results of model comparison from the Bayesian evidence are discussed in Sect.\ \ref{sec:mcmc}

%

\end{document}